\documentclass{aa}

\usepackage{graphicx}
\usepackage{txfonts}
\usepackage[dvipsnames]{xcolor}
\usepackage{natbib}
\usepackage{longtable}
\usepackage{siunitx}
\usepackage{multirow}
\usepackage{hyperref}
\usepackage[normalem]{ulem}

\hypersetup{
  colorlinks= true,   
  urlcolor  = blue,     
  linkcolor = blue, 
  citecolor = blue
}

\makeatletter
\renewcommand*\aa@pageof{, page \thepage{} of \pageref*{LastPage}}
\makeatother

\bibpunct{(}{)}{;}{a}{}{,}    

\newcommand{\meta}{12+\log\left(\mathrm{O/H}\right)}
\newcommand{\eqmeta}{$12+\log\left(\mathrm{O/H}\right)$}

\defcitealias{Kobulnicky2004}{KK04}
\defcitealias{Tremonti2004}{T04}
\defcitealias{Pilyugin2016}{P16}
\defcitealias{Pettini2004}{PP04}
\defcitealias{Dopita2016}{D16}

\usepackage{orcidlink}

\begin{document}

\title{The Lockman--SpReSO project}

\subtitle{Main properties of infrared selected star-forming galaxies}

\author{Mauro González-Otero \inst{1,2,3}\orcidlink{0000-0002-4837-1615}     
      \and
      Jordi Cepa \inst{1,2,3} \orcidlink{0000-0002-6566-724X}
      \and
      Carmen P. Padilla-Torres \inst{1,2,3,4} \orcidlink{0000-0001-5475-165X}
      \and
      Maritza A. Lara-López \inst{5}\orcidlink{0000-0001-7327-3489}
      \and
      J. Jesús González \inst{6}\orcidlink{0000-0002-3724-1583}
      \and
      Ángel Bongiovanni \inst{7,3}\orcidlink{0000-0002-3557-3234}
      \and
      Bernabé Cedrés \inst{7,3}\orcidlink{0000-0002-7382-6407}
      \and
      Miguel Cerviño \inst{8}\orcidlink{0000-0001-8009-231X} 
      \and
      Irene Cruz-González \inst{6}\orcidlink{0000-0002-2653-1120}
      \and
      Mauricio Elías-Chávez \inst{6}\orcidlink{0000-0002-0115-8374}
      \and
      Martín Herrera-Endoqui \inst{6}\orcidlink{0000-0002-8653-020X}
      \and
      Héctor J. Ibarra-Medel \inst{6}\orcidlink{0000-0002-9790-6313}
      \and
      Yair Krongold \inst{6}\orcidlink{0000-0001-6291-5239}
      \and
      Jakub Nadolny \inst{9}\orcidlink{0000-0003-1440-9061}
      \and
      C. Alenka Negrete \inst{6}\orcidlink{0000-0002-1656-827X}
      \and
      Ana María Pérez García \inst{8,3} \orcidlink{0000-0003-1634-3588}
      \and
      José A. de Diego \inst{6} \orcidlink{0000-0001-7040-069X}
      \and
      J. Ignacio González-Serrano \inst{10,3} \orcidlink{0000-0003-0795-3026}
      \and
      Héctor Hernádez-Toledo \inst{6}\orcidlink{0000-0001-9601-7779}
      \and
      Ricardo Pérez-Martínez \inst{11,3}\orcidlink{0000-0002-9127-5522}
      \and
      Miguel Sánchez-Portal \inst{7,3}\orcidlink{0000-0003-0981-9664}
      }

\institute{Instituto de Astrofísica de Canarias, 
        E-38205 La Laguna, 
        Tenerife, Spain 
    \and Departamento de Astrofísica, Universidad de La Laguna (ULL), 
        E-38205 La Laguna, Tenerife, 
        Spain 
    \and Asociación Astrofísica para la Promoción de la Investigación, Instrumentación y su Desarrollo, ASPID, 
        E-38205 La Laguna, 
        Tenerife, Spain 
    \and Fundación Galileo Galilei-INAF Rambla José Ana Fernández Pérez, 7,  
        E-38712 Breña Baja, 
        Tenerife, Spain 
    \and Departamento de Física de la Tierra y Astrofísica, Instituto de Física de Partículas y del Cosmos, IPARCOS. Universidad Complutense de Madrid (UCM), 
        E-28040, 
        Madrid, Spain. 
    \and Instituto de Astronomía, Universidad Nacional Autónoma de México, 
        Apdo. Postal 70-264, 04510 
        Ciudad de México, Mexico 
    \and Institut de Radioastronomie Millimétrique (IRAM), Av. Divina Pastora 7, Núcleo Central 
        E-18012, 
        Granada, Spain 
    \and Centro de Astrobiología (CSIC/INTA), 
        E-28692 ESAC Campus, Villanueva de la Cañada, 
        Madrid, Spain 
    \and Astronomical Observatory Institute, Faculty of Physics, Adam Mickiewicz University, ul.~S{\l}oneczna 36, 
        60-286 Pozna{\'n}, Poland 
    \and Instituto de Física de Cantabria (CSIC-Universidad de Cantabria), 
        E-39005, 
        Santander, Spain 
    \and ISDEFE for European Space Astronomy Centre (ESAC)/ESA, P.O. Box 78, 
        E-28690 Villanueva de la Cañada, 
        Madrid, Spain 
    \\
    \email{mauro.gonzalez-ext@iac.es, mauromarago@gmail.com}
}        
\date{Received ---; accepted ---}
    
 
  \abstract
   {}
   {In this article we perform a comprehensive study using galaxy data from the Lockman--SpReSO project, a far-infrared (FIR) selected sample of galaxies observed using optical spectroscopy. We analysed a sub-sample of star-forming galaxies (SFGs) with secure spectroscopic redshifts, mostly in the luminous infrared galaxies domain. From these galaxies, parameters such as extinction, star formation rate (SFR), and metallicity were derived. The present paper examines how these properties evolve in relation to each other, and in comparison with low-redshift FIR and non-FIR-selected samples of galaxies.}
   {We applied distinct selection criteria to attain an SFG sample with minimal AGN contamination. Multiple approaches were used to estimate the intrinsic extinction, SFR and gas-phase metallicity for the SFGs. In conjunction with findings in the literature, we examined the correlation between SFRs and stellar masses ($M_*$), as well as the metallicity evolution depending on $M_*$. Finally, the 3D relationship between $M_*$, SFR and metallicity, is also studied.}
   {From the initial spectroscopic sample of 409 FIR-selected objects from the Lockman--SpReSO catalogue, 69 (17\%) AGNs have been identified and excluded, which is nearly double the percentage found in local studies, leaving a sample of 340 SFGs. The analysis of the $M_*$--SFR relationship revealed that Lockman--SpReSO IR-selected SFGs show signs of evolution at redshifts $z>0.4$, shifting above the main sequence, with a mean value of $\sim0.4$ dex. They are located within the starburst galaxy region since 78\% of the galaxies fall into this category. In addition, no evident flattening was found in the relation to specific SFR with redshift for $\log M_* (M_\odot) \gtrsim 10.5$. In line with the $M_*$--metallicity relation (MZR) outcomes published in previous studies for optically selected SFGs, however, during the analysis of the MZR, it was found that IR-selected SFGs exhibit lower metallicities than those anticipated on the basis of their $M_*$ and redshift. During the investigation of the 3D $M_*$--SFR--metallicity relation (FP), it was established that the research sample is consistent with relations in the existing literature, with an average scatter of  $\sim0.2$ dex. However, a re-calibration of the FP when using the SFR obtained from the IR luminosity is required and, in this case, no attenuation in the correlation for $\log M_* (M_\odot) \gtrsim 10.5$ is observed. This result points to a possible evolution of the more massive fraction of the sample in the sense of decreasing the present-day star formation with respect to the averaged star formation in the past.}
   {}


    \maketitle

\section{Introduction} \label{sec:1}
Studying the evolution of galaxies is challenging, since it involves possible variations with redshift of relationships involving global indicators, such as star formation rate (SFR), metallicity (Z), stellar mass ($M_*$) and other related parameters. Estimating these indicators, each with its specific uncertainties and intrinsic limitations, is a difficult task, as is shown later in this paper.

Extensive research on the evolution of the main sequence (MS), comparing star formation rate (SFR) and stellar mass ($M_*$), has been undertaken. Numerous studies, such as those of \cite{Brinchmann2004,Speagle2014,Popesso23}, have investigated this indicator, including \cite{Cedres2021}, who found no evolution of this indicator, even for low-mass galaxies below $z\simeq 1.43$.

The mass--metallicity relationship (MZR), which reflects the enrichment of galactic gas compared to the mass within stars, serves as an additional observational indicator of evolution \citep[][and references therein]{Duarte2022}. The results so far obtained indicate that metallicity rises with $M_*$ and cosmic time \citep[for example,][]{Sanders2021}, and is inversely correlated with SFR, as evidenced by the fundamental mass--metallicity--SFR (FP or FMR) relationship \citep{Lara2010,Mannucci2010}. The MZR definition suggests that gas accretion, outflows and metal astration probably influence it, but it could also be affected by factors such as downsizing or infrared (IR) luminosity. However, its possible evolution remains uncertain. For instance, at the lower redshift ($z\sim0.4$) and low-mass end $\left(\log\left(M_*\right) < 8\right)$, \cite{Nadolny2020} found no evidence of MZR evolution. This was also confirmed at higher redshifts up to $z = 2.3$ by \cite{Cresci2019} and up to $z =3.3$ by \cite{Sanders2021}, including the low-mass end. However, according to \cite{Pistis2022}, the MZR is subject to biases resulting from S/N ratio and quality flags in the spectra, leading to overestimated metallicities or to the selection of high-metallicity galaxies. These authors, however, observed that the relationship between metallicity and specific SFR (sSFR) is relatively insensitive to such biases. Nevertheless, \cite{Henry2021} found evolution in MZR and FMR at redshifts $1.3 < z < 2.3$ using a larger sample than those of previous authors, and extended to low mass galaxies. In addition, these authors confirmed this evolution for a sub-sample of galaxies with high S/N spectra.

The situation described above is even more complex for far-IR (FIR) selected galaxies. Luminous infrared galaxies (LIRGs) and ultra-luminous infrared galaxies (ULIRGs) are galaxies with total infrared luminosities ($L_{\rm TIR}$, from 8 to 1000 $\mu$m) between 10$^{11}$--10$^{12}$ and 10$^{12}$--10$^{13}$ solar luminosities (L$_\odot$), respectively. Both LIRGs and ULIRGs are considered interacting/merging or post-merger galaxies  \citep[see for example,][and references therein]{Kilerci2014,Pereira2019,Nadolny2023}. From a sample of nine (U)LIRGs at redshifts $0.2 < z < 0.4$, \cite{Pereira2019} concluded that 10--25\% are isolated discs and the rest interacting or merging systems, with SFR(H$\alpha$) $\sim$ SFR($L_{\rm TIR}$) but with interstellar medium (ISM) conditions different from those in local galaxies. From a sample of 20 LIRGs at low--intermediate redshifts ($0.25<z<0.65$) that were classified to be in the regime between normal and starburst galaxies, however, \cite{Lee2017} concluded that only 10\% show signs of interaction. From a sample of 118 local ULIRGs with mean redshift $z\simeq$0.18 selected from SDSS DR10, \cite{Kilerci2014} found that SFR(H$\alpha$) is in mean 8 times lower that SFR($L_{\rm TIR}$), and that $Z$ determined from optical lines using R$_{23}$ (see Section \ref{sec:meta_theor}) is on average about 0.3 dex lower with respect to local SDSS galaxies. From a study of five local ULIRGs, however, \cite{Chartab2022} claim that the lower metallicity observed in ULIRGs is an artefact originating from metallicity determinations using optical instead of FIR lines. Nevertheless, determining metallicity using FIR lines remains controversial. \cite{Herrera2018} still reports lower metallicity in (U)LIRGs, even when using FIR lines. This is in contrast to \cite{Chartab2022}, who used different lines.

The objective of this paper is to study the relations among different determinations of SFR, $M_*$ and metallicity for a statistically significant sample of star-forming galaxies(SFGs) at intermediate redshifts, selected according to their IR emission and having a robust redshift determination. The study will further analyse possible differences with respect to optically selected samples, other LIRG-selected samples and their possible evolution.

This paper is organized as follows. In Section \ref{sec:2} the data available for this work are presented. The different methods of discriminating SFGs with respect to active galactic nuclei (AGN) are described in Section \ref{sec:3}. The extinction correction adopted is explained in Section \ref{sec:4}. Sections \ref{sec:5} and \ref{sec:6} explain the different gas metallicity and SFR estimators, respectively. Section \ref{sec:7} presents the results of the global indicators MS, MZR and FP. Finally, our conclusions are given in Section \ref{sec:8}. Throughout the paper, magnitudes are given in the AB system \citep{Oke1983}. The cosmological parameters adopted are: $\Omega_\mathrm{M} = 0.3$, $\Omega_\mathrm{\Lambda} = 0.7$, and $H_\mathrm{0} = 70$ km s$^{-1}$ Mpc$^{-1}$. We assume a \cite{Chabrier2003} initial mass function (IMF) for the estimation of both SFR and $M_{*}$.


\section{Data selection} \label{sec:2}
\begin{figure*}
    \includegraphics[trim= 0cm 0cm 0cm 0cm,width=\textwidth]{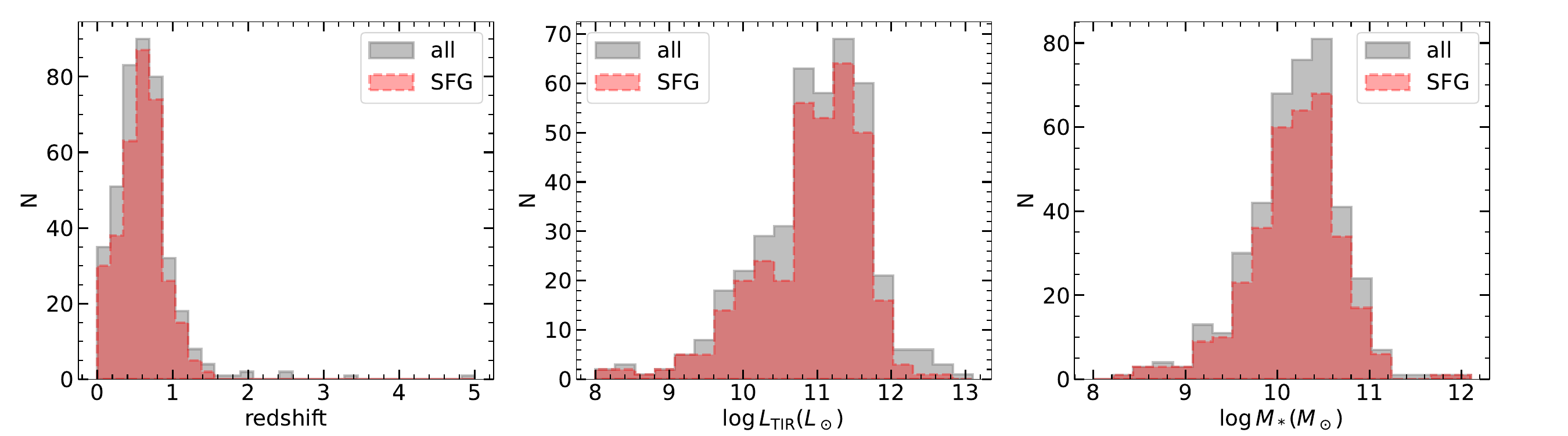}
    \caption{Distributions of the spectroscopic redshift, $L_\mathrm{TIR}$ and $M_*$. In grey the sample of 409 objects collected from the Lockman--SpReSO catalogue with a determined spectroscopic redshift is represented. The objects selected as SFG are represented in red (see Sect. \ref{sec:3} for details).}
    \label{fig:prop_distr}
\end{figure*}

The data used in this paper are drawn from the Lockman--SpReSO project described in \cite{Gonzalez2022}, to which the reader is referred for a detailed description of the observations, reduction and catalogue compilation. Further details on the optical and FIR wavelength coverage, fluxes, optical spectral resolution, area covered and ancillary data available are also explained in \cite{Gonzalez2022}.

In summary, the Lockman--SpReSO project involves an optical spectroscopic follow-up of 956 sources selected by their FIR  flux using \textit{Herschel Space Observatory} data. In addition, 188 objects of interest, with a limiting magnitude in the Cousins $R$ band ($R_{\rm C}$) of $R_{\rm C}<24.5$ mag and all located in the Lockman Hole field were included in the sample. The spectroscopic observations were conducted using the WHT/A2F-WYFFOS\footnote{\url{https://www.ing.iac.es/Astronomy/instruments/af2}} \citep{wyffos2014} and WYIN/HYDRA\footnote{\url{https://www.wiyn.org/Instruments/wiynhydra.html}} instruments for objects within the brighter subset of the catalogue ($R_{\rm C}<20.6$ mag). While for objects in the fainter subset ($R_{\rm C}>20$ mag), GTC/OSIRIS\footnote{\url{http://www.gtc.iac.es/instruments/osiris/osiris.php}} \citep{JCepa2000} was used.

Of the 1144 sources of Lockman--SpReSO, spectroscopic analysis allowed the determination of spectroscopic redshift for 456 objects, where 357 come from objects with at least two identified spectral lines and 99 were obtained using only one spectral line secured using all available photometric information. To ensure a robust determination, we have used the available photometric bands information, generally from FUV to FIR \citep[see appendix A in][]{Gonzalez2022}, the photometric redshifts available in the literature, and the intensity and shape of the spectral line. In this paper we analyse the 456 regardless of how the redshift was obtained. Furthermore, with the spectroscopic redshifts obtained, \cite{Gonzalez2022} conducted spectral energy distribution (SED) fitting using the available photometric data spanning from the ultraviolet to the FIR wavelength range. This was carried out using the CIGALE software (Code Investigating GALaxy Emission, \citealt{Bugarella2005}, \citealt{Boquien2019}). In particular, this SED fitting method provides more accurate measurements of $M_{*}$ and $L_{\rm TIR}$ than previous determinations also derived from SED fittings but based on photometric redshifts.

The Lockman--SpReSO project catalogue includes a set of sources that were not selected based on their infrared emission. These objects were added to complement the observational masks and are also of interest. The selection comprises radio galaxies, obscured quasars, and distant galaxies that were initially thought to be X-ray binaries and cataclysmic star candidates \citep[for more details see][]{Gonzalez2022} Out of the 456 objects with determined spectroscopic redshift in the catalogue, 47 belong to this sample of non IR selected objects. Therefore, for the purpose of this work, they must be removed from the studied sample.

Thus, for the development of this work we selected 409 objects for which the spectroscopic redshift had been determined, spanning the range $0.03 \lesssim z \lesssim 4.96$ with a median redshift of 0.6.  In this sample, 54\% of the sources are LIRGs, 6\% ULIRGs and 1\% hyper-luminous infrared galaxies (HLIRGs, $L_\mathrm{TIR}> 10^{13}$ $L_\odot$) with a median value $\log L_\mathrm{TIR}(L_\odot) = 11.1$. The $M_*$ of this sample lies in the range $8.23 \lesssim \log M_* (M_\odot) \lesssim 12.1$ with a median value $\log M_* (M_\odot)=10.26$. The distributions of these properties are shown in Fig.\ \ref{fig:prop_distr}, where the sample of 409 objects is represented by the grey distribution. In addition, a signal-to-noise ratio (S/N) greater than 3 was applied to all spectroscopic lines used in the subsequent sections.


\section{Star-forming galaxies and AGN discrimination} \label{sec:3}

The sample selected in the previous section did not differentiate between SFGs and AGNs. However, for the study proposed in this article, it is essential to distinguish between AGNs and SFGs in order to calculate accurately the extinction, SFR and metallicity. The investigation of the AGN population in the Lockman--SpReSO project will be presented in separate papers scheduled for imminent publication.

To classify the objects as AGNs or SFGs, we used a combination of photometric, spectroscopic, and SED fitting data gathered from \cite{Gonzalez2022}, as described below. Photometric criteria, based on X-ray and IR information of the objects, were utilized, along with spectroscopic criteria, which involved analyzing spectral lines to perform this classification.

\subsection{Photometric criteria}
\subsubsection{X-ray discrimination criteria}

The use of X-ray data to differentiate between SFG and active galactic nuclei is common practice. The strong X-ray emission from the accretion disc regions surrounding the central black holes serves as a robust indicator of the nature of the objects. One of the initial studies in this area was conducted by \cite{Maccacaro1988}, who used the ratio of X-ray-to-optical flux (X/O ratio) as a means of distinguishing AGN from other sources of X-ray emission. Subsequent studies have also used the X/O ratio to differentiate AGN from other X-ray sources \citep{Stocke1991,Lehmann2001,Szokoly2004,Xue2011,Luo2017,Chen2018,Marina2019,Elias2021}.

In our study, we adopted the criterion described by \cite{Luo2017}, who conducted research within the spectroscopic redshift range of $0 \lesssim z \lesssim 5$. In their work the X/O ratio is given by:
$$\log \mathrm{X/O}  = \log F_\mathrm{X}  + 0.4\, R_\mathrm{C} + 4.77 > -1,$$
where $F_\mathrm{X}$ is the X-ray flux within the range 0.2--12 keV and $R_\mathrm{C}$ is the magnitude in the $R_\mathrm{C}$ band, which is used as a tracer for the optical flux. The X-ray data of the sample were obtained from observations made by the \textit{XMM-Newton} space telescope over the Lockman field and the 4XMM-DR10 catalogue \citep{Webb2022}.

\begin{figure*}
    \includegraphics[trim= 0cm 0cm 4.5cm 0cm,width=\textwidth]{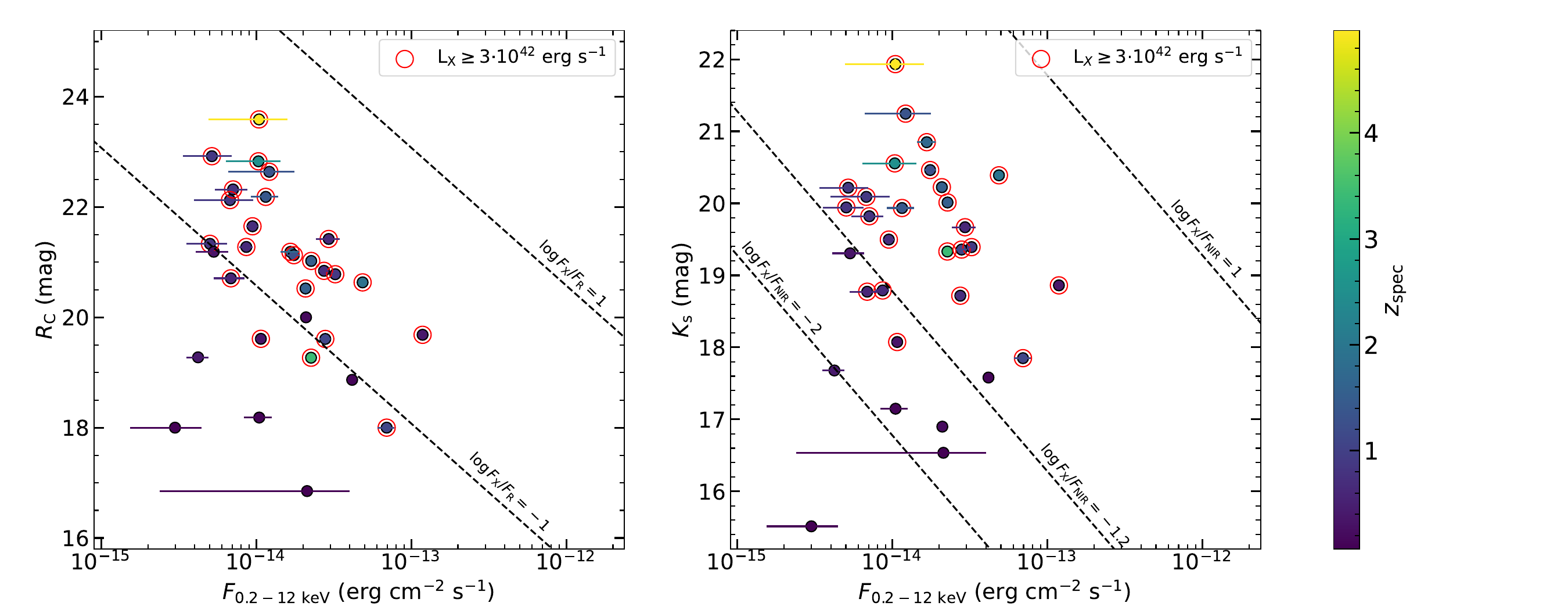}
    \caption{AGN classification based on X/O and X/NIR ratios. The left panel illustrates the relationship between X-ray flux (0.2 to 12 keV) and $R_C$ magnitude. Dashed lines represent the thresholds for $\log\left(\mathrm{X/O}\right)>-1$ and $\log\left(\mathrm{X/O}\right)>1$, with objects above $\log\left(\mathrm{X/O}\right)>-1$ classified as AGN. Red circles highlight points meeting the AGN criterion $L_\mathrm{X} > 3\times10^{42}$ erg s$^{-1}$ of \cite{Xue2011}. Colour coding indicates the spectroscopic redshift of the objects. The right panel displays the X-ray flux plotted against $K_s$ magnitude. Objects above the  $\log\left(\mathrm{X/NIR}\right)>-1.2$ threshold are categorised as AGN. Red circles and colour coding remain consistent with the left panel.}
    \label{fig:xray_class}
\end{figure*}

The left panel of Fig. \ref{fig:xray_class} shows the X-ray flux versus the $R_\mathrm{C}$ magnitude. The dashed lines mark the regions where $\log \mathrm{X/O} >-1$ and $\log \mathrm{X/O} >1$. Using the above criterion, we have classified objects above the $\log \mathrm{X/O} >-1$ threshold as AGN. The colour code in the figure represents the spectroscopic redshift of the objects. This diagnostic diagram classified 21 objects as AGN.

The ratio between the X-ray flux and the near-IR flux (X/NIR ratio), using the $K_\mathrm{s}$ band as an indicator of the NIR \citep{Luo2017}, could also be used to separate AGN from SFGs. The criterion for this separation is defined as:
$$\log \mathrm{X/NIR} = \log F_\mathrm{X} + 0.4\, K_\mathrm{s} + 5.29 > -1.2$$
The outcome after applying this criterion is shown in the right panel of Fig.\ \ref{fig:xray_class}, where the X-ray flux is plotted against the $K_\mathrm{s}$ magnitude. The dashed lines represent the limits in the X/NIR ratio, and the colour code in the figure represents the spectroscopic redshift. Following the definition in \cite{Luo2017}, the objects in the range $\log \mathrm{X/NIR} >-1.2$ were categorised as AGN. A total of 21 objects were classified as AGN using this criterion.

The final criterion applied using X-ray information is that defined by \cite{Xue2011}, who defined a threshold for the X-ray luminosity ($L_\mathrm{X}$) to distinguish AGN from other X-ray sources. According to this criterion, any source with $L_\mathrm{X} \geq 3\times10^{42}$ erg/s is classified as an AGN. In Fig. \ref{fig:xray_class} this criterion is represented by red circles over the points. The application of this criterion resulted in the classification of 24 objects as AGN. A total of 25 unique objects were classified as AGN by at least one of the above criteria based on X-ray information.

\subsubsection{Infrared discrimination criteria}
One of the characteristics of AGN that helps us to distinguish them from SFGs is the fact that they tend to be redder in the NIR and MIR. This is because the SED of AGN from the UV down to $\sim5$ $\mu$m is usually dominated by a power-law continuum, whereas SFGs show a black-body continuum with a peak above $\sim1.6$ $\mu$m due to the underlying stellar population \citep{Stern2005}.

\begin{figure}
    \includegraphics[trim= 0cm 0cm 0cm 0cm,width=\linewidth]{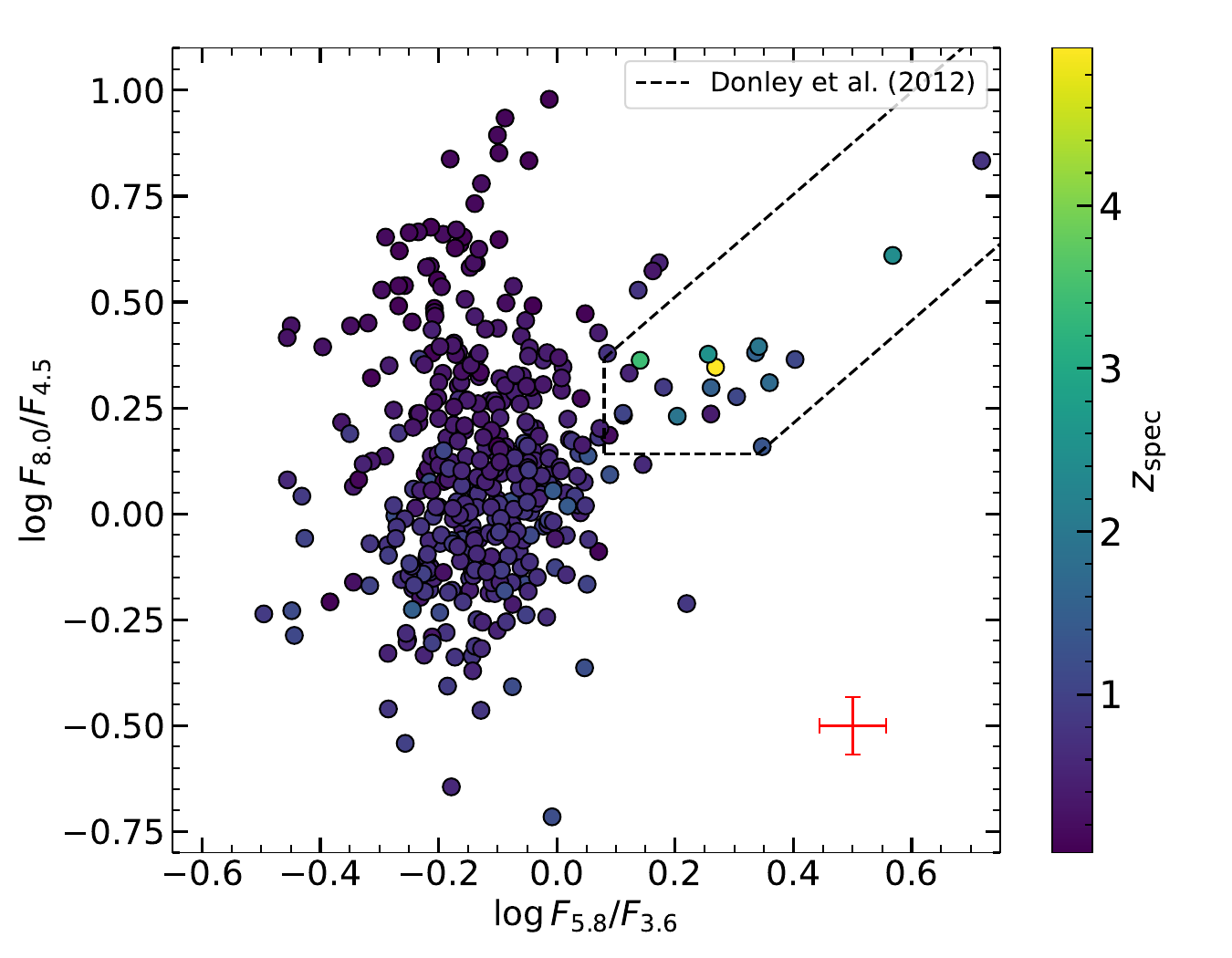}
    \caption{Criteria for separating AGN from SFGs, using \textit{Spitzer}/IRAC bands, updated by \cite{Donley2012}. The x-axis represents the ratio of fluxes in the 5.8 $\mu$m and 3.6 $\mu$m bands, while the y-axis represents the ratio of fluxes in the 8.0 $\mu$m and 4.5 $\mu$m bands. The area bounded by the black dashed lines corresponds to the selection criterion defined by \cite{Donley2012}. The colour coding represents the spectroscopic redshift of the objects. The average $1 \sigma$ size is shown in red at the bottom right.}
    \label{fig:donley}
\end{figure}

Using the above information, investigations have been carried out using the IR information to separate AGN from SFGs. \cite{Donley2012} used the four \textit{Spitzer}/IRAC bands (3.6, 4.5, 5.8, and 8.0 $\mu$m) to distinguish AGN from SFGs, by updating the \cite{Lacy2004} and \cite{Stern2005} criteria, which suffer from contamination by normal SFGs in deep IRAC data. They defined an empirical region where AGN lie in the \textit{Spitzer}/IRAC colour space:

$$x\geq 0.08; \, y \geq 0.15$$
$$ y \geq 1.21 \, x - 0.27$$
$$ y \leq 1.21 \, x + 0.27$$
$$f_{4.5\, \mu m} > f_{3.6\, \mu m}; \, f_{5.8\, \mu m} > f_{4.5\, \mu m}; \, f_{8.0\, \mu m}>f_{5.8\, \mu m}$$
where $x=\log\left( f_{5.8\, \mu m}/f_{3.6\, \mu m}\right)$ and $y =\log\left(f_{8.0\, \mu m}/f_{4.5\, \mu m}\right)$. The definition of this criterion is independent of the redshift within the sample under study. In Fig. \ref{fig:donley} we plot the above flux ratios and the region defined by \cite{Donley2012} where the objects are classified as AGN. Based on the results of this diagnostic diagram, 19 objects were classified as AGN.

\begin{figure}
    \includegraphics[trim= 0cm 0cm 0cm 0cm,width=\linewidth]{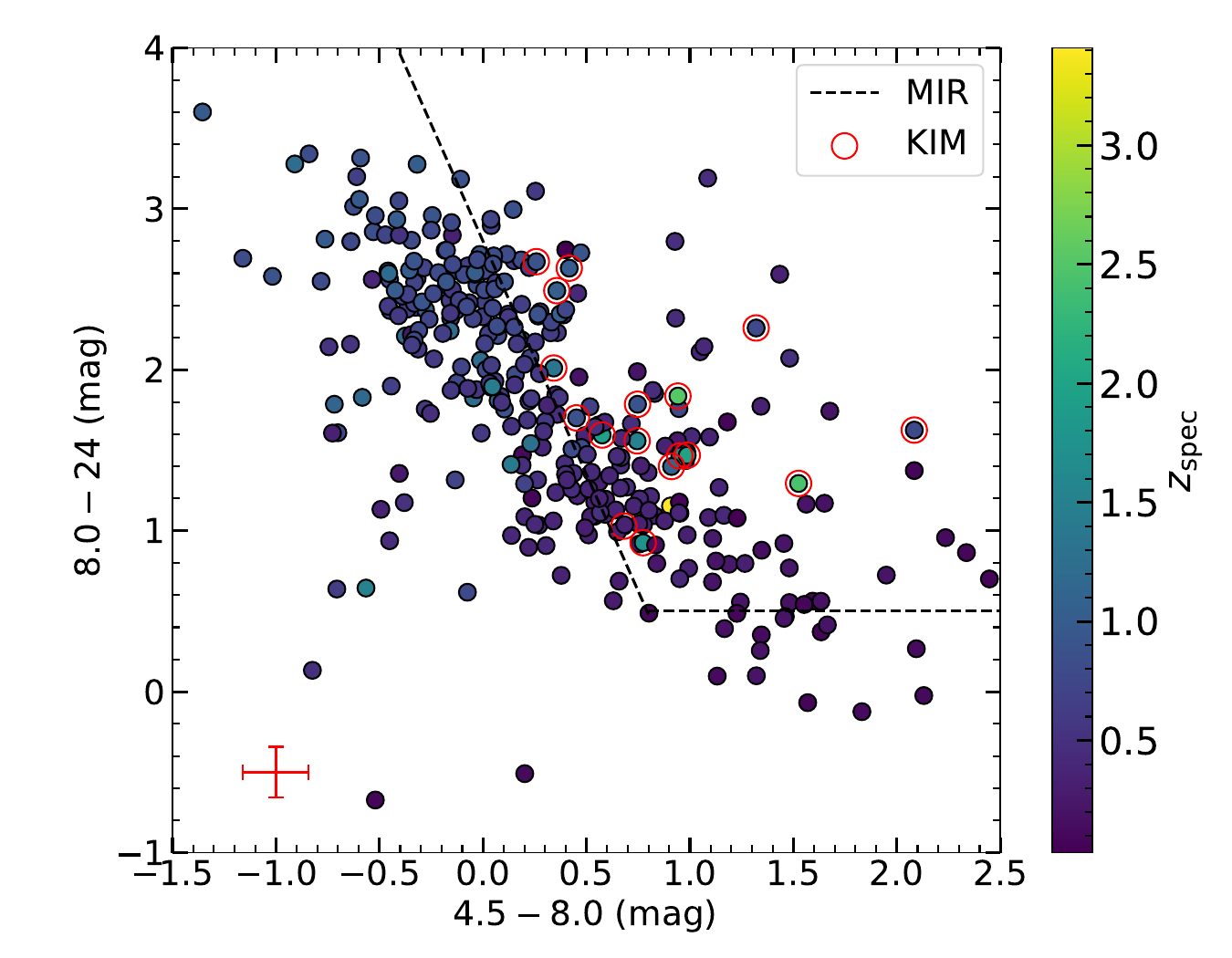}
    \caption{Separation criteria from \cite{Messias2012} using the \textit{Spitzer}/IRAC 4.5 and 8.0 $\mu$m bands, the \textit{Spitzer}/MIPS 24 $\mu$m band, and the $K_\mathrm{s}$ band. The dashed line outlines the area where AGN could be found based on the IM criteria. The red empty circles indicate the objects that also met the KIM criteria. The colour coding represents the spectroscopic redshift of the objects. The average $1 \sigma$ size is shown in red at the bottom left.}
    \label{fig:MIR_KIM}
\end{figure}

The criterion developed by \cite{Messias2012} has been further applied to the sample. This criterion is a classification method that uses information from the \textit{Spitzer}/IRAC 4.5 and 8.0 $\mu$m bands, and the 24 $\mu$m \textit{Spitzer}/MIPS band to define a region where the AGN would be found, called the IRAC-MIPS criterion (``IM'' criterion), with an additional criterion based on the $K_\mathrm{s}$ band (``KIM'' criterion):
$$ [8.0] - [24] > -2.9 \, \left(\left[4.5\right] - \left[8.0\right]\right) + 2.8$$
$$ [8.0] - [24] >  0.5 $$
$$ K_\mathrm{s} - [4.5] > 0 $$
where [4.5], [8.0], [24], and $K_\mathrm{s}$ are the AB magnitudes in the 4.5 and 8.0 $\mu$m IRAC bands, the 24 $\mu$m MIPS band, and the $K_\mathrm{s}$ band, respectively. This criterion minimizes the contamination of the selected AGN sample by normal and SFGs at low redshifts thanks to the addition of the criterion using the $K_\mathrm{s}$ band, while it strongly separates SFGs from AGN at high redshifts. The fact that it is independent of redshift fits perfectly with the Lockman--SpReSO data, since the sample is not constrained by redshift. In Fig.\ \ref{fig:MIR_KIM} we plot the colour between the 4.5 and 8.0 $\mu$m bands against the colour between the 8.0 and 24 $\mu$m bands. The dashed line represents the area defined by the IM criterion and the objects with and empty red circle are those that also satisfied the KIM criterion. Thus, using the \cite{Messias2012} criterion, a total of 17 objects were classified as AGN. This leaves a total of 26 unique objects classified as AGN using the IR photometric discrimination criteria.

\subsection{Spectroscopic criteria}
Spectral emission lines emanating from intricate interactions between photons and ionized atoms serve as diagnostics of the ionization sources, chemical composition and physical conditions within galaxies. In the context of distinguishing between AGN and SFGs, the precise identification of ionization mechanisms becomes particularly crucial.

One of the most widely used diagrams for this purpose, based on spectral emission lines, is the well-known  Baldwin, Phillips \& Terlevich (BPT) diagram \citep{BPT1981}. This diagram uses the ratios of optical emission lines to distinguish between SFGs and AGN, where the most common ratios are [\ion{O}{III}]$\lambda$5007/\ion{H}{$\beta$} and [\ion{N}{II}]$\lambda$6584/\ion{H}{$\alpha$}. AGN exhibit a higher [\ion{O}{III}]$\lambda$5007/\ion{H}{$\beta$} ratio owing to intense radiation from the accretion disc whereas SFGs have a lower [\ion{O}{III}]$\lambda$5007/\ion{H}{$\beta$} ratio, as their emission lines are mainly influenced by ionization from young stars. Similarly, the [\ion{N}{II}]$\lambda$6584/\ion{H}{$\alpha$} ratio is higher in AGN compared to SFGs, a result of stronger emission lines from the ionized gas around the black hole, leading to distinctive regions occupied by AGN and SFGs in this diagram.

\begin{figure}
    \includegraphics[trim= 0cm 0cm 0cm 0cm,width=\linewidth]{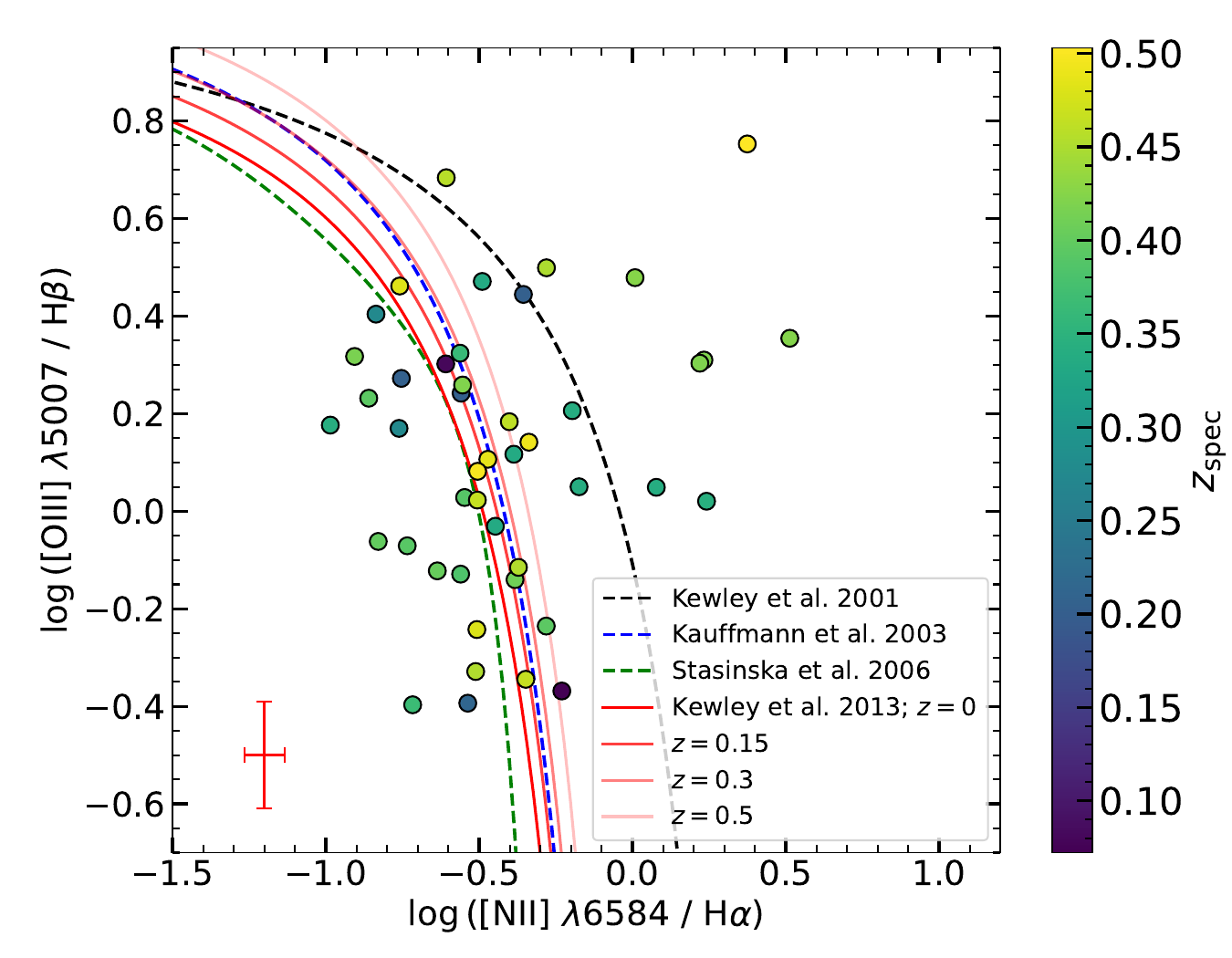}
    \caption{Representation of the BPT diagram defined by \cite{BPT1981} to separate SFGs from AGN. The orange, blue and black dashed lines represent the \cite{Stasinska2006}, \cite{Kauffmann2003} and \cite{Kewley2001} selection criteria, respectively. The red solid lines represent the \cite{Kewley2013} selection criteria for redshifts 0, 0.15, 0.3, and 0.5. The colour coding represents the spectroscopic redshift of the objects. The average $1 \sigma$ size is shown in red at the bottom left.}
    \label{fig:BPT}
\end{figure}

Figure \ref{fig:BPT} shows the BPT diagram for the sample of objects from Lockman--SpReSO used in this paper. The criteria from \cite{Kewley2001}, \cite{Kauffmann2003}, \cite{Stasinska2006}, and \cite{Kewley2013} to separate SFGs from AGN are shown. The \cite{Kewley2001} criterion is the least restrictive for SFGs, allowing composite galaxies to be included in the selection. Criteria such as those of \cite{Stasinska2006} or \cite{Kauffmann2003} are more restrictive and filter out SFGs more efficiently. We decided to use the \cite{Kewley2013} selection criterion, since it takes into account the evolution of the line ratios with redshift. This is especially important for our sample, which extends up to a $z\sim0.5$, and the \cite{Kewley2013} criterion allowed us to better separate SFGs from AGN. According to this diagnostic diagram, 26 objects were classified as SFGs and 19 objects were classified as AGN or composite galaxies.

\begin{figure}
    \includegraphics[trim= 0cm 0cm 0cm 0cm,width=\linewidth]{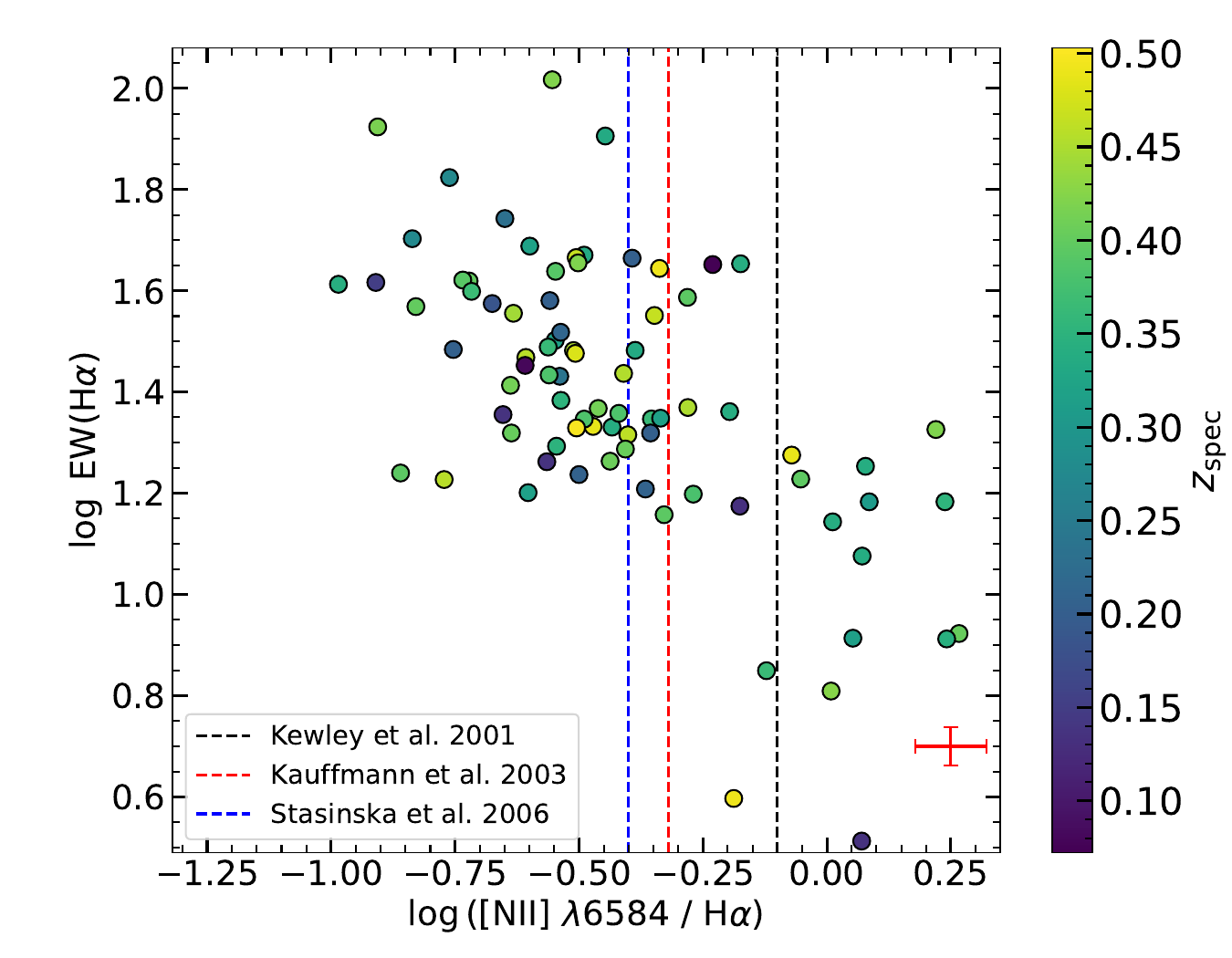}
    \caption{EW$\alpha$n2 criterion defined by \cite{CidFernandes2010} for the classification of SFGs and AGN. The \cite{Stasinska2006} criterion (blue dashed line) has been adopted for the separation, as it is the most restrictive for SFGs. The \citet[][red dashed line]{Kauffmann2003} and \citet[][black dashed lines]{Kewley2006} criteria are also shown. The colour coding represents the spectroscopic redshift of the objects. The average $1 \sigma$ size is shown in red at the bottom right.}
    \label{fig:CidFernandes}
\end{figure}

As a complement to the BPT diagram, the classification diagram developed by \cite{CidFernandes2010}, named the EW$\alpha$n2 diagram, is very valuable for objects with a limited number of emission lines available. This method uses only the \ion{H}{$\alpha$} line and the [$\ion{N}{II}$] line. Using these two lines, a degeneracy appears between Seyfert and AGNs which the authors solved by adding the \ion{H}{$\alpha$} rest-frame equivalent width (EW). The separation between SFGs and AGNs is established by criteria based on the [$\ion{N}{II}$] / \ion{H}{$\alpha$} ratio \citep{Kewley2001,Kauffmann2003,Stasinska2006}.
In this case, we adopted the more restrictive criteria of \cite{Stasinska2006}, which define SFGs to occupy the region with $\log\ [$\ion{N}{II}$] / $\ion{H}{$\alpha$} $\leq-0.4$, and AGN are defined to be in the region with $\log\ [$\ion{N}{II}$] / $\ion{H}{$\alpha$} $\geq -0.2$. Figure \ref{fig:CidFernandes} shows the sample objects on the EW$\alpha$n2 diagram, where the different separation criteria mentioned above have been marked. The application of this diagnostic diagram resulted in the classification of 31 objects as not being SFGs.

In summary, compiling all the results, we have obtained 25 objects classified as AGN using X-ray-based criteria, 26 using IR-based criteria, and an additional 33 AGN based on spectroscopic criteria. This yields a total of 69 unique objects classified as not SFGs from the sample of 409 objects taken from the Lockman--SpReSO catalogue, representing 17\% of the sample. This value is slightly higher than the values of 11.4\% and 11.5\% found by \cite{Lara2013} for the SDSS and GAMA surveys, respectively. However, this result is highly dependent on the selection criteria of the initial sample. \cite{Sabater2019} revealed that 20\% of their radio-galaxy sample with a counterpart in SDSS were AGN, whilst \cite{MAgliocchetti2018} found for the VLA-COSMOS catalogue that 33\% of the galaxies were AGN. The spectroscopic redshift, that spans the range $0.03\lesssim z\lesssim 1.52$ with a median value $\sim0.6$, $L_\mathrm{TIR}$, and $M_*$ distributions of the SFG selected are shown in the Fig.\ \ref{fig:prop_distr}, where they are marked in red.

\section{Extinction correction}\label{sec:4}
In the study of galaxies, spectral lines serve as invaluable tools, providing crucial information about their physical properties, chemical composition, and ionization processes. Nonetheless, accurate interpretation of spectral lines can be significantly hindered by the presence of extinction effects caused by interstellar dust within the galaxies themselves. Extinction occurs when dust particles in the interstellar medium absorb and scatter light, leading to a reddening of the observed spectra. This reddening effect can introduce systematic biases in the measurements of emission lines, potentially misleading the derived physical parameters of galaxies, such as SFR and metallicities.

Correcting for extinction effects therefore becomes of paramount importance in obtaining reliable and precise measurements of emission lines. In this section, we present our methodology for correcting extinction in the spectral lines of galaxies.

\subsection{Stellar absorption of underlying older components}
Before tackling the task of extinction correction, we need to consider the contribution of the old stellar population to line measurements. One of the properties of this type of stellar population is absorption in the Balmer series lines, which is often superimposed on the emission lines produced by the excitation of gas by the hotter, younger stars.

To correct for this effect, we have adopted the criterion of \cite{Hopkins2003,Hopkins2013}, using a constant value of 2.5 \AA\ for the EW correction, EW$_{\rm c}$, to account for the absorption contribution arising from the underlying stellar population. It is important to note that this form of correction is recommended primarily for cases where it is desired to study the properties of a large sample of objects as a whole. For detailed analyses of individual objects, more refined measurements of the underlying absorption should be preferred.

Correction of the Balmer line fluxes for the effect of the underlying stellar absorption has been performed using:
\begin{equation}
  F = F_{\rm obs} \, \left(1 + \frac{\mathrm{EW}_\mathrm{c}}{\rm EW} \right) ,
\end{equation}
where $F$ is the underlying absorption-corrected flux, $F_{\rm obs}$ is the observed flux of a Balmer line, $\mathrm{EW}_\mathrm{c}$ is the applied correction of 2.5 \AA, and EW is the equivalent width of the line.

\subsection{Extinction calculation}

To perform the extinction correction on the emission lines of SFGs, we have made use of the empirical relationships established by \cite{Calzetti1994}, who state that the intrinsic flux ($F_{\rm int}$) at a given wavelength ($\lambda$) can be obtained as follows
\begin{equation}\label{eq:flux_corr}
    F_{\rm int}(\lambda)= F(\lambda) \, 10^{0.4\, A(\lambda)} = F(\lambda) \, 10^{0.4\, k(\lambda)\, E(B-V)},
\end{equation}
where $F(\lambda)$ is the observed flux at that wavelength corrected for the underlying stellar absorption, $A(\lambda)$ is the extinction at that wavelength, and $k(\lambda)$ is the reddening curve evaluated at that wavelength. In this work we have used the reddening curve defined by \cite{Calzetti2000}. Finally, $E(B-V)$ is the colour excess; that is, the variation that the $B-V$ colour suffers due to the effect of the dust.

The estimation of the extinction from the Balmer decrement of the emission lines observed in the spectra of the objects is one of the most reliable methods because, for a fixed electronic temperature, quantum physics gives the theoretical values for the ratios of the lines. Thus, the colour excess is obtained using the H$\alpha/$H$\beta$ ratio:
\begin{equation}
    E(B-V) = \frac{2.5}{k(\lambda_{\mathrm{H}\beta})-k(\lambda_{\mathrm{H}\alpha})}\,\log_{10} \left( \frac{\left(\mathrm{H}\alpha/\mathrm{H}\beta\right)_{\rm obs}}{\left(\mathrm{H}\alpha/\mathrm{H}\beta\right)_{\rm th}} \right),
\end{equation}
where $k(\lambda_{\mathrm{H}\beta})$ and $k(\lambda_{\mathrm{H}\alpha})$ are the values of the reddening curve evaluated at H$\beta$ and H$\alpha$ wavelengths, respectively; $\left(\mathrm{H}\alpha/\mathrm{H}\beta\right)_{\rm obs}$ is the observed Balmer decrement, and $\left(\mathrm{H}\alpha/\mathrm{H}\beta\right)_{\rm th}$ is the quantum-physical value of the Balmer decrement in the case of non-extinction. The standard case adopted in the study of SFGs is the recombination Case B described by \cite{Osterbrock1989}, where $\left(\mathrm{H}\alpha/\mathrm{H}\beta\right)_{\rm th} = 2.86$ is defined for an electron temperature $T= 10^4\,\mathrm{K}$ and an electron density $n_e = 10^2\,\mathrm{cm}^{-3}$.

The sample of SFGs has a redshift distribution with a median value of $\sim0.6$, which means that more than half of the sample does not have the H$\alpha$ line available because it is outside the spectral range covered for objects at redshifts $\gtrsim0.5$. Other orders of the Balmer decrement are proposed to calculate the extinction; for example, the ratio $\mathrm{H}\beta/\mathrm{H}\gamma$. The theoretical value for this ratio, defined under the recombination Case B of \cite{Osterbrock1989}, is $\left(\mathrm{H}\beta/\mathrm{H}\gamma\right)_{\rm th}=2.13$. Subsequent orders of the Balmer decrement were not considered because the lines are weaker and usually have a low S/N ratio.

\begin{figure*}
    \includegraphics[trim= 0cm 0cm 0cm 0cm,width=\textwidth]{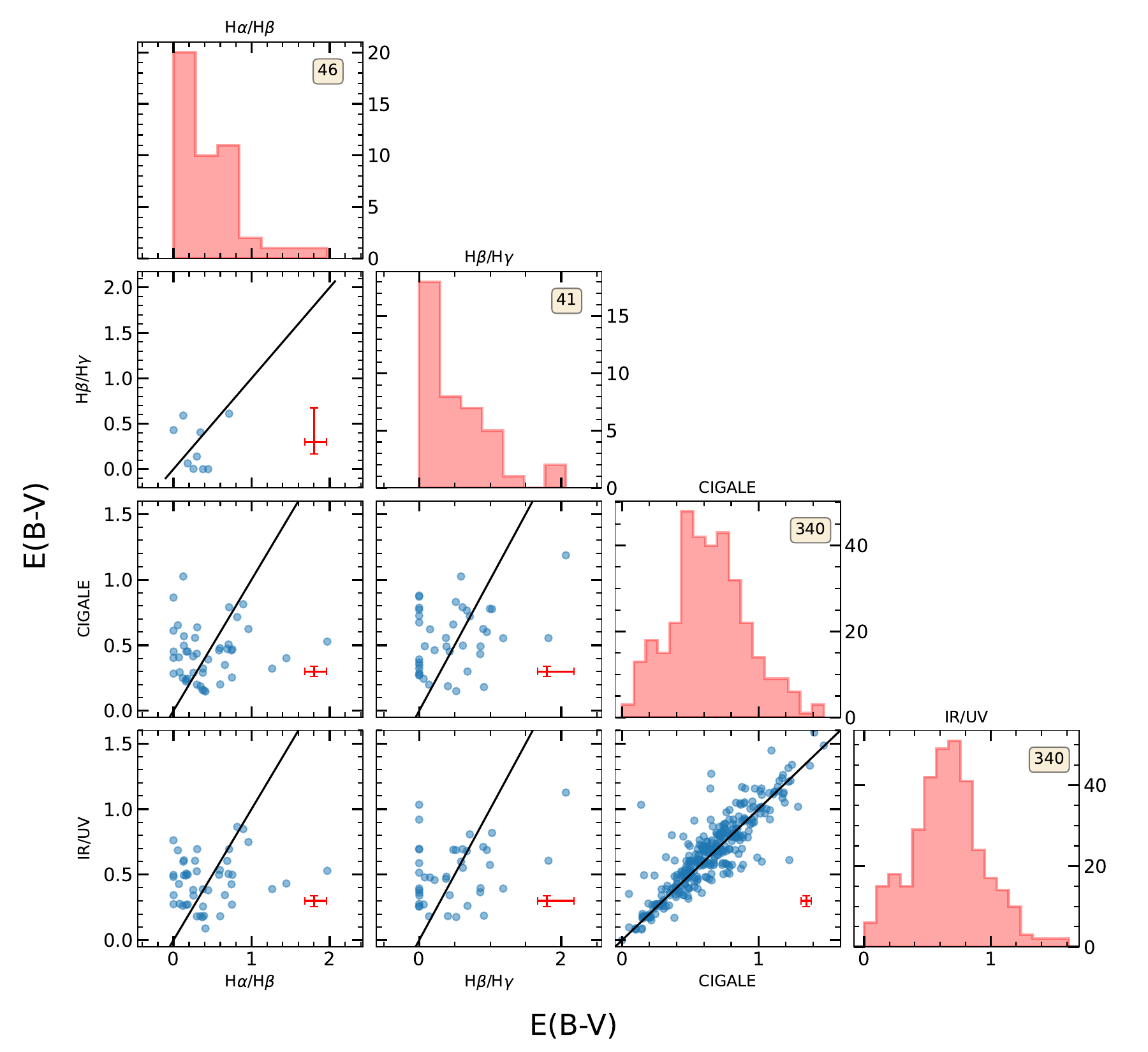}
   \caption{Histograms and comparisons of the colour excesses obtained using the tracers analysed in this article. The black lines indicate the relationship $x=y$. The values in the inset boxes refer to the number of objects for which the colour excess could be calculated with the respective tracer. The average $1 \sigma$ size is shown in red at the bottom right of each panel.}
    \label{fig:EBV}
\end{figure*}

Figure \ref{fig:EBV} shows the distributions and the relation for the $E(B-V)$ obtained using the $\mathrm{H}\alpha/\mathrm{H}\beta$ and $\mathrm{H}\beta/\mathrm{H}\gamma$ ratios. The number of sources for which the $E(B-V)$ can be obtained is limited. Therefore, we have considered other ways of calculating $E(B-V)$ for the extinction correction.

From the SED fits performed by \cite{Gonzalez2022} using CIGALE (the CIGALE configuration is depicted in their appendix B), the $E(B-V)$ of the nebular lines was obtained for each source. CIGALE also performs a parameter determination by Bayesian inference, taking into account all the models with which an attempt has been made to fit the SED of an object. This value obtained from Bayesian inference is the one we have taken as $E(B-V)$. The relation with the other tracers and the distribution obtained for the $E(B-V)$ provided by the SED fits using CIGALE can be seen in Figure \ref{fig:EBV}.

The ultimate method to measure $E(B-V)$ is based on the IR/UV ratio. By investigating the balance between the IR and UV wavelengths, one can gain an understanding of the extinction phenomenon as the dust absorbs the UV radiation from hot stars and re-emits it in the IR spectrum. We adopt the parameterisation method developed by \cite{CaiNaHao2011} to determine the colour excess, and this is accomplished by using the IR/UV ratio. This method provides a colour excess for the continuum and for comparing it with the one obtained previously, a commonly used conversion factor is applied:
\begin{equation}
  E(B-V)_{\rm c} = 0.44\times E(B-V),
\end{equation}
where $E(B-V)_{\rm c}$ represents the colour excess of the continuum. Figure \ref{fig:EBV} shows the distribution obtained and the comparison with the other tracers studied. The $E(B-V)$ value obtained from the SED fits with CIGALE is the one used to correct for the extinction of the line fluxes (Eq. \ref{eq:flux_corr}) used in the following sections of the paper. Moreover, it is in good agreement with the IR/UV tracer, as shown in Fig. \ref{fig:EBV}.

\section{Gas phase metallicities estimation}\label{sec:5}

\begin{figure*}
    \includegraphics[trim= 0cm .5cm 0cm 0cm,width=\textwidth]{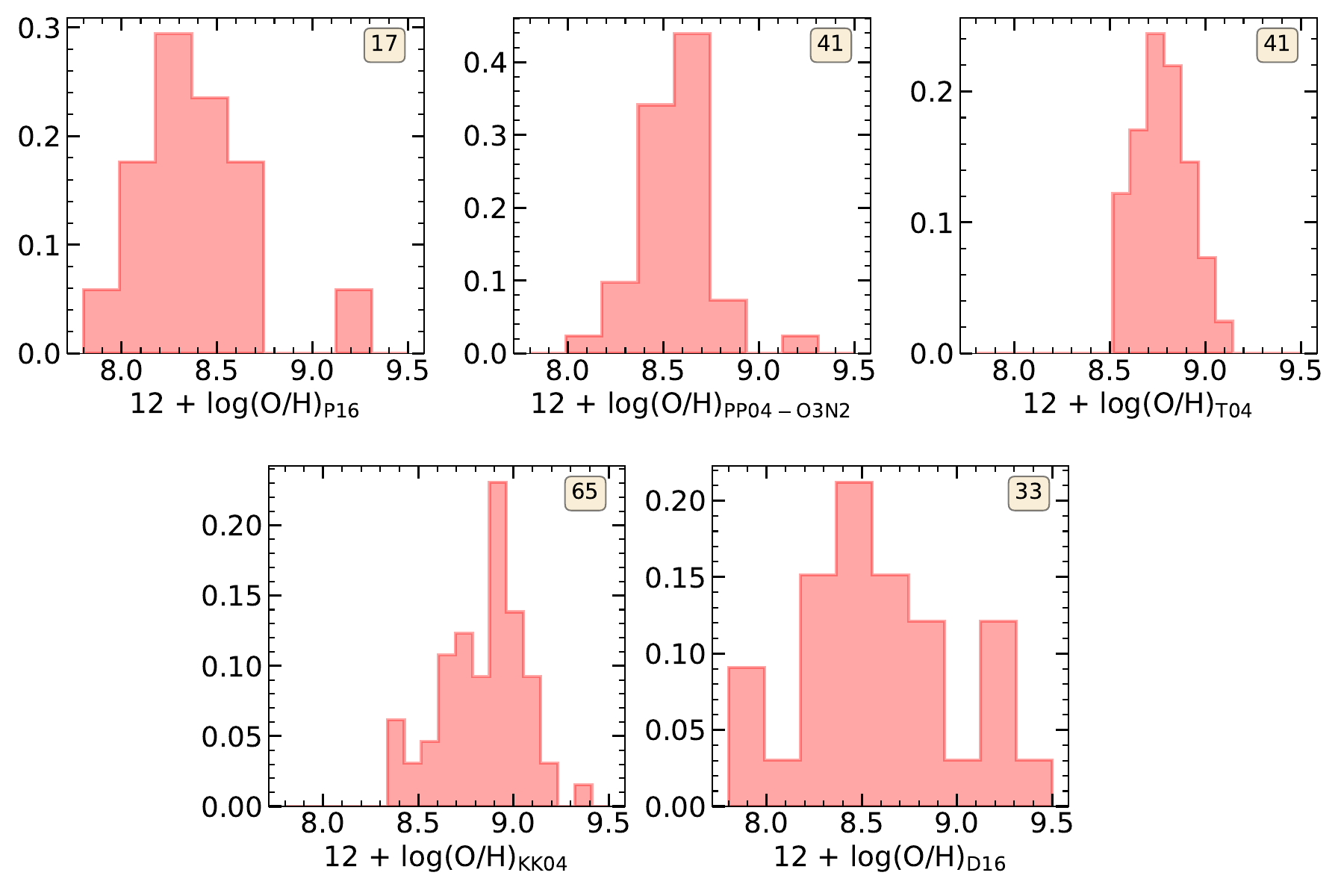}
    \caption{Normalized histograms of the metallicities obtained in this article for the SFGs of the Lockman--SpReSO project. From left to right and from top to bottom, the parameterisations are from \cite{Pilyugin2016}, \cite{Pettini2004}, \cite{Tremonti2004}, \cite{Kobulnicky2004}, and \cite{Dopita2016}. The values in the inset boxes refer to the number of objects for which the gas-phase metallicity could be calculated with the respective tracer.}
    \label{fig:meta_hist}
\end{figure*}

In this section, we examine the calibrations used to determine the gas-phase metallicity of the SFGs in our sample. This determination is based on the oxygen abundance, \eqmeta.

The direct method of estimating the electron temperature ($T_e$) of the ionized gas requires achieving high resolution and S/N ratio. However, this method uses weak auroral lines such as [\ion{O}{III}] $\lambda4363$ or [\ion{N}{II}] $\lambda5755$, which are difficult to observe, so other methods have been proposed. Some of them are empirical calibrations of the $T_e$ method, while others are based on photoionisation models. This variety of methods leads to a lack of universality in the metallicity calibrator. In addition, the discrepancies introduced by the variations between methods lead to deviations in the fundamental metallicity relation and its associated projections, adding complexity to the study.

\subsection{Empirical methods}

One of the empirical methods analysed in this study is the \citet[hereafter P16]{Pilyugin2016} calibration, which derives the abundance of oxygen using the intensities of strong emission lines in \ion{H}{II} regions. They separated the calculation of the metallicity for the upper and lower branches characteristic of the methods based on oxygen determination. The upper branch is thus defined as  $\log\left(N_2\right)<-0.6$, where $N_2 = [\ion{N}{II}]\, \lambda\lambda6548,84/\mathrm{H}\beta$. The metallicity equation is
\begin{multline}
    \meta = 8.589 + 0.022\, \log\left(R_3/R_2\right) + 0.399\,\log\left(N_2\right)  \\ 
    + \left(-0.137+0.164\,\log\left(R_3/R_2\right) + 0.589\,\log\left(N_2\right)\right)\,\log\left(R_2\right).
\end{multline}
The lower branch definition is $\log\left(N_2\right)>-0.6$ and the metallicity equation is:
\begin{multline}
    \meta = 7.932 + 0.944\, \log\left(R_3/R_2\right) + 0.695\,\log\left(N_2\right) \\ 
    + \left(0.970-0.291\,\log\left(R_3/R_2\right) + 0.19\,\log\left(N_2\right)\right)\,\log R_2.
\end{multline}
In both equations the coefficients are $R_2=[\ion{O}{II}]\,\lambda\lambda3727,29/\mathrm{H}\beta$ and $R_3 = [\ion{O}{III}]\,\lambda\lambda4959,5007/\mathrm{H}\beta$. The distribution of the metallicity obtained is plotted in the left panel in the upper row in Fig. \ref{fig:meta_hist}.

The relationship derived by \citet[hereafter PP04]{Pettini2004} is the second empirical tracer we studied based on the O3N2 estimator. It is particularly useful because it can be applied to high-redshift galaxies, as it employs spectral lines that are in close proximity to one another, eliminating the need for complex procedures such as extinction corrections or flux calibrations, which are challenging when observing high-redshift galaxies in the infrared. Their parameterisation of the metallicity is:
\begin{equation}
    \meta = 8.73 - 0.32\times \mathrm{O3N2} ,
\end{equation}
where $\mathrm{O3N2} = \log \left[ \left(\left[\ion{O}{III}\right]\lambda5007/\mathrm{H}\beta\right)/\left(\left[\ion{N}{II}\right]\lambda6584/\mathrm{H}\alpha\right) \right]$ and is valid only for galaxies with $\mathrm{O3N2}<2$. The result is plotted in the middle panel of the upper row in Fig. \ref{fig:meta_hist}.

\subsection{Theoretical photoionisation models-based methods}\label{sec:meta_theor}
\citet[hereafter T04]{Tremonti2004} developed an objective calibration of oxygen abundance by fitting the most intense emission lines in the optical range with theoretical model approaches. The fitting models were created by combining single stellar population (SSP) models from \cite{Bruzual2003} with photoionisation models from CLOUDY \citep{Ferland1998}. The parameterisation of the oxygen abundance is based on the $R_{23}$ estimator, where $R_{23} = \left(\left[\ion{O}{II}\right]\lambda\lambda3727,29+\left[\ion{O}{III}\right]\lambda\lambda4959,5007\right)/\mathrm{H}\beta$. The metallicity is calculated using the following equation:
\begin{equation}
    \meta = 9.185 - 0.313\,x - 0.264\,x^2 - 0.321\,x^3    
\end{equation}
where $x = \log\left( R_{23}\right)$ and is valid only for the upper branch of the double-valued $R_{23}$-abundance relation. The previous definition of the upper branch is not applicable to objects with redshift $z\gtrsim0.45$, as the [\ion{N}{II}] lines lie outside the Lockman--SpReSO spectra. To differentiate between the upper and lower branches, we have set the criteria defined by the region $\log R_{23} >0.85$ and $\log M_*<9.3$ for the upper branch. The motivation for these criteria can be found in Appendix \ref{sec:appendixA}. The right panel in the upper row in Fig.\ \ref{fig:meta_hist} shows the metallicity distribution obtained.

The metallicity estimates derived from the $R_{23}$ estimator, enable the calculation of metallicity for a larger number of Lockman--SpReSO objects, owing to the use of shorter wavelength spectral lines that are still visible in the optical, for objects at higher redshifts.

The \citet[hereafter KK04]{Kobulnicky2004} parameterisation is an iterative technique for determining the oxygen abundance that also relies on the $R_{23}$ estimator. The $R_{23}$ calibrator is sensitive to the ionization state of the gas, characterized by the ionization parameter ($q$), which is the number of hydrogren-ionizing photons passing through a unit area per second divided by the hydrogen density of the gas. The ionization parameter is determined through the [\ion{O}{II}]/[\ion{O}{III}] ratio of lines, which is in turn influenced by the metallicity of the gas, through the subsequent equation:
\begin{multline}
     \log q = \left[ 32.81-1.153 \, y^2     \right. \\ 
     \left. \qquad + \left[ \meta \right] \left( -3.396 -0.025 \, y + 0.1444 \, y^2 \right) \right]\  \\
      \qquad \times \left[ 4.603 -0.3119 \, y - 0.163 \, y^2 \right. \\
     \left. \qquad  + \left[\meta\right]\left( -0.48 + 0.0271\, y + 0.02037 \, y^2 \right)\right]^{-1},
\end{multline}
where $y = \log \left( \left[\ion{O}{III}\right] \lambda 5007/ \left[\ion{O}{II}\right]\lambda 3727\right)$. An initial metallicity value is required for the ionization parameter calculation. To determine it, we analyse which branch the object belongs to, according to the already established criteria, and assign an initial value of \eqmeta = 8.2 for the lower branch and \eqmeta = 8.7 for the upper branch. The obtained ionization parameter value is then used to determine the metallicity via the following parameterisation:
\begin{multline}
    \meta_{\rm lower} = 9.40 + 4.65\,x - 3.17\,x^2 \\
    -\left( 0.272 + 0.547\,x -0.513\,x^2\right)\log q,
\end{multline}
\begin{multline}
    \meta_{\rm upper} = 9.72 - 0.777\,x - 0.951\,x^2 - 0.072\,x^3 \\
    \qquad \qquad \qquad - 0.811\,x^4 -\left( 0.0737 - 0.0713\,x - 0.141\,x^2 \right. \\
    \left. + 0.0373\,x^3 - 0.058\,x^4\right) \log q,
\end{multline}
where $x=\log R_{23}$, and upper and lower sub-indices indicate the branch. This process is repeated until $\meta$ converges. The obtained distribution is shown in the left panel in the bottom row of Fig. \ref{fig:meta_hist}.

Finally, we implemented the \citet[hereafter D16]{Dopita2016} criterion to determine the oxygen abundance. This methodology utilizes lines which are redder than those used in the previously mentioned techniques, namely H$\alpha$, [\ion{N}{II}]$\lambda$6484 and the [\ion{S}{II}] $\lambda\lambda$ 6717,31 doublet. These lines are also similar in wavelength, hence extinction correction can be neglected. The computation of metallicity is thus established by the following relation:
\begin{equation}
    \meta = 8.77 + y + 0.45\,(y+0.3)^5,
\end{equation}
where $y = \log\left( [\ion{N}{II}]/[\ion{S}{II}]\right) + 0.264\,\log\left([\ion{N}{II}]/\mathrm{H}\alpha\right)$. The left panel in the bottom row in Fig. \ref{fig:meta_hist} shows the distribution obtained.


\section{Star Formation rate}\label{sec:6}

\begin{figure*}
    \includegraphics[trim= 0cm 0cm 0cm 0cm,width=\textwidth]{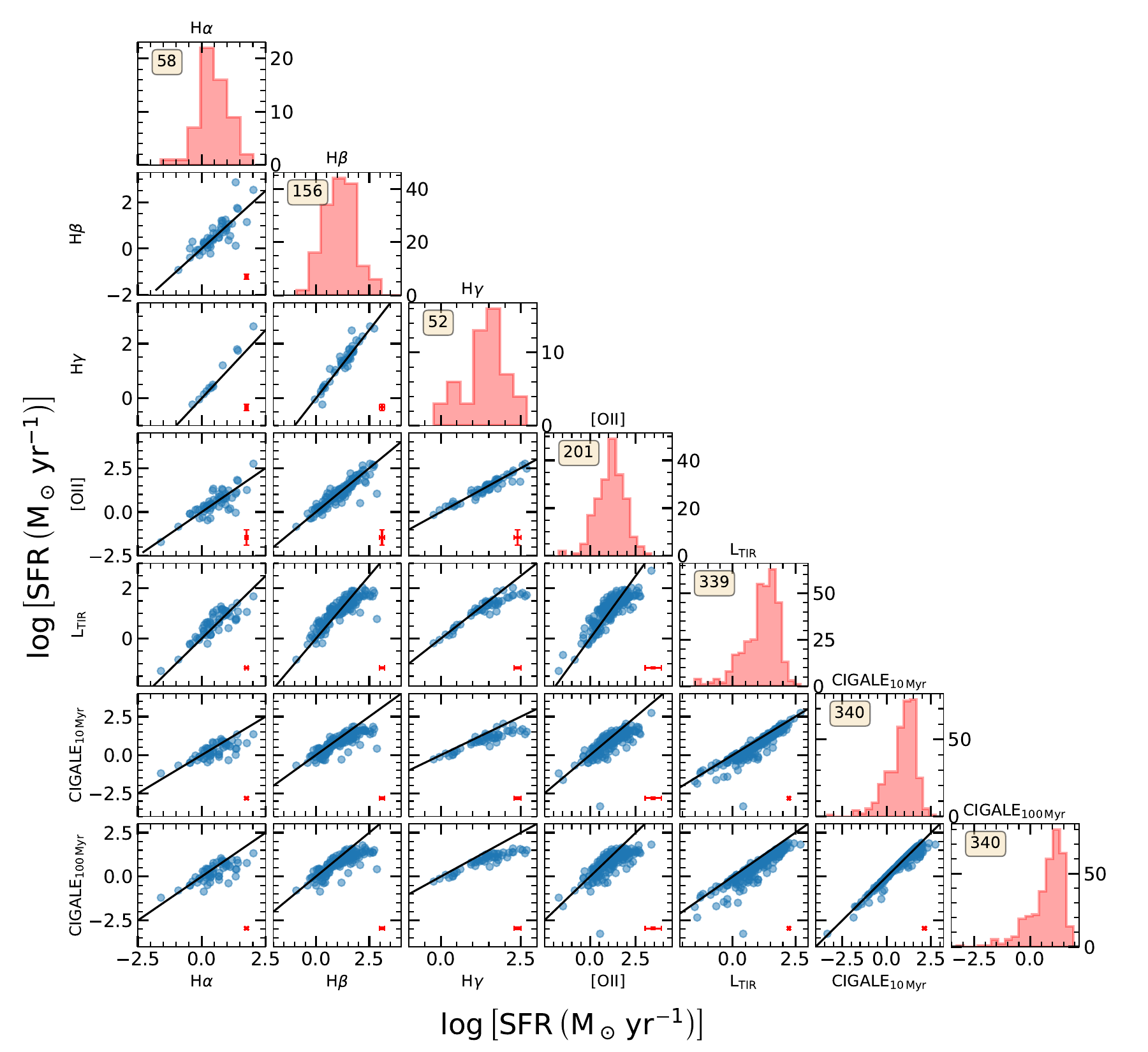}
   \caption{Histograms and comparisons of the SFR obtained using different traces for the SFG from the Lockman--SpReSO project. The values in the histogram panels indicate the total number  for which the SFR could be calculated with the respective tracer. The solid black lines represent the $x=y$ relation. The values in the inset boxes refer to the number of objects for which the SFR could be calculated with the respective tracer. The average $1 \sigma$ size is shown in red at the bottom right of each panel.}
    \label{fig:SFR_corner}
\end{figure*}

\subsection{Spectral lines}
The H$\alpha$ line is the primary SFR indicator in the optical range of the local universe. The emission of the H$\alpha$ line originates from the \ion{H}{II} regions, wherein massive newly formed stars ionize the gas, resulting in the production of Balmer and other emission lines. The H$\alpha$ emission is moreover uninfluenced by the metallicity of the gas or the star formation history. There are many calibrations that use the H$\alpha$ flux to determine the SFR. One of the most commonly used methods, and the one adopted in this paper, is the calibration proposed by \cite{Kennicutt2012}:
\begin{equation}
    \log\left[\mathrm{SFR} \left( M_\odot\,\mathrm{yr}^{-1}\right)\right] = \log\left[L_\mathrm{H\alpha} \left(\mathrm{erg}\,\mathrm{s^{-1}}\right)\right] - 41.27.
\end{equation}
However, the H$\alpha$ line falls outside the optical spectrum for objects with redshifts $z\gtrsim0.5$. The majority of objects in the Lockman--SpReSO sample have higher redshifts, so we have to rely on other spectral lines to calculate the SFR. The H$\beta$ line, available for objects up to redshift $z\sim1$, can be used as a tracer of SFR. Assuming a relation with the H$\alpha$ line, typically under the recombination Case B of \cite{Osterbrock1989}, the same used for extinction correction, the SFR derived using H$\beta$ is:
\begin{multline}
    \log\left[\mathrm{SFR} \left( M_\odot\,\mathrm{yr}^{-1}\right)\right] = \\
    \log\left[L_\mathrm{H\beta} \left(\mathrm{erg}\,\mathrm{s^{-1}}\right)\right] - 41.27 + \log 2.86,
\end{multline}
where the 2.86 factor is the theoretical value for the $\mathrm{H}\alpha/\mathrm{H}\beta$ ratio in the supposed recombination Case B. 

Under the same considerations, the SFR can be determined from the H$\gamma$ line flux. This line is observable for objects with a redshift of up to $z\sim1.3$; however, it is comparatively fainter, more affected by extinction and less detectable in the Lockman--SpReSO spectra. The equation for determining the SFR is:
\begin{multline}
    \log\left[\mathrm{SFR} \left( M_\odot\,\mathrm{yr}^{-1}\right)\right] = \\
    \log\left[L_\mathrm{H\gamma} \left(\mathrm{erg}\,\mathrm{s^{-1}}\right)\right] - 41.27 + \log 2.86 + \log 2.13,
\end{multline}
where 2.13 is the theoretical values for the  $\mathrm{H}\beta/\mathrm{H}\gamma$ ratio in the supposed recombination Case B.

Another spectral line that can be utilized to determine the SFR is the $\left[\ion{O}{II}\right] \lambda\lambda 3726,29$ doublet, which appears in the same regions as H$\alpha$ and represents similar star formation timescales. However, it is less correlated with the emission created by the ionization of gas from massive stars. On the other hand, extinction in the region where this doublet is located is significant and depends greatly on the metallicity and the ionization parameter. Nevertheless, there are parameterisations that use the $\left[\ion{O}{II}\right]$ flux to determine the SFR with good results. For this research, we have adopted the parameterisation obtained by \cite{Figueira2022}, who took into account the metallicity of their SFG sample for their study. The equation is:
\begin{equation}
    \log\left[\mathrm{SFR} \left( M_\odot\,\mathrm{yr}^{-1}\right)\right] = 0.96 \, \log\left[L_\mathrm{\left[\ion{O}{II}\right]} \left(\mathrm{erg}\,\mathrm{s^{-1}}\right)\right] -39.69.
\end{equation}

\subsection{CIGALE data products}
Indicators based on photometric luminosity in selected bands are also used as tracers for SFR. One of the most commonly used is $L_{\rm TIR}$, based on energy balance studies: UV emission from the hottest stars, absorbed by dust, is re-emitted in the IR regime of the electromagnetic spectrum. It should be noted that the timescale of the SFR studied using L$_{\rm TIR}$ is greater ($\sim100\,\mathrm{Myr}$) than that derived from the optical spectral lines ($\sim10\,\mathrm{Myr}$).

In this paper we use the relationship between SFR and $L_{\rm TIR}$ described by \cite{Kennicutt2012}. The mathematical expression is:
\begin{equation}
    \log\left[\mathrm{SFR} \left( M_\odot\,\mathrm{yr}^{-1}\right)\right] = \log\left[L_{\rm TIR} \left(\mathrm{erg\,s^{-1}}\right)\right] - 43.41,
\end{equation}
where $L_{\rm TIR}$ was obtained from the paper of \cite{Gonzalez2022}. They performed a SED-fitting process on the Lockman--SpReSO objects, from which, among other parameters, $L_{\rm TIR}$ was obtained.

Among the results given by CIGALE there is also an estimate of the SFR. CIGALE provides the estimate of the instantaneous SFR and the SFR averaged over 10 and $100\,\mathrm{Myr}$. For comparison with the results obtained with the spectral lines and the $L_{\rm TIR}$, we have selected the SFR averaged over 10 and $100\,\mathrm{Myr}$, respectively.

Figure \ref{fig:SFR_corner} shows the results obtained for the SFR with the different methods analysed in this section. The histograms represent the distributions obtained by each method, while the plots show the relationships between the different tracers. As we have already mentioned, we need to bear in mind that the SFR derived from spectral lines study very similar time ranges (0--10 Myr), whereas the SFR derived from $L_{\rm TIR}$ studies longer times (0--100 Myr), so the comparison between these tracers is purely indicative.

Table \ref{tab:sample_catalogue} presents an excerpt from the Lockman--SpReSO SFG catalogue. The table displays the quantities obtained in this work for the galaxy properties studied in Sects. \ref{sec:4}, \ref{sec:5}, and \ref{sec:6}.

\begin{sidewaystable*}
  \centering
  \caption{Sample of the values derived in this study for the properties of the SFGs in the Lockman--SpReSO catalogue.}
  \label{tab:sample_catalogue}
  \begin{tabular}{ccccccccccc}
  \hline \hline 
  objid & $z_\mathrm{spec}$ & $\log M_* (M_\odot)$ & $\log L_\mathrm{TIR} (L_\odot)$ & $E(B-V)_{H\alpha/H\beta}$ & $E(B-V)_{H\beta/H\gamma}$ & $E(B-V)_\mathrm{CIGALE}$ & $E(B-V)_\mathrm{IR/UV}$ & $\meta_{\rm P16}$ & $\meta_{\rm PP04}$ \\ [0.4ex] 
  (1)   &  (2)              &  (3)                 & (4) & (5) & (6) & (7) & (8) & (9) & (10) \\ [0.4ex] 
  \hline 
96702  & 0.54950$\pm$ 0.00005 & 10.00 $\pm$ 0.13 & 10.92 $\pm$ 0.12 & - & - & $0.62_{-0.08}^{+0.09}$ & $0.67_{-0.11}^{+0.11}$ & - & - \\  [0.6ex]
99389  & 0.81001$\pm$ 0.00005 & 10.59 $\pm$ 0.08 & 11.64 $\pm$ 0.05 & - & - & $0.88_{-0.04}^{+0.04}$ & $0.91_{-0.12}^{+0.14}$ & - & - \\  [0.6ex]
132869 & 0.41953$\pm$ 0.00008 & 10.48 $\pm$ 0.04 & 10.84 $\pm$ 0.03 & - & - & $0.40_{-0.03}^{+0.03}$ & $0.40_{-0.08}^{+0.07}$ & - & - \\  [0.6ex]
94458  & 0.67084$\pm$ 0.00004 & 10.77 $\pm$ 0.08 & 11.73 $\pm$ 0.02 & - & - & $0.76_{-0.02}^{+0.02}$ & $0.89_{-0.11}^{+0.09}$ & - & - \\  [0.6ex]
123801 & 0.61025$\pm$ 0.00006 & 10.65 $\pm$ 0.09 & 11.12 $\pm$ 0.06 & - & - & $0.75_{-0.08}^{+0.08}$ & $0.72_{-0.12}^{+0.10}$ & - & - \\  [0.6ex]
  \hline
  \end{tabular}

\bigskip \bigskip
\renewcommand\thetable{1} 
\caption{\textit{continued}}

\begin{tabular}{ccccccccccc}
  \hline \hline 
  objid & $\meta_{\rm T04}$ & $\meta_{\rm KK04}$ & $\meta_{\rm D16}$ & $\log[\mathrm{SFR}_{\rm H\alpha} (M_\odot\mathrm{yr}^{-1})]$ & $\log[\mathrm{SFR}_{\rm H\beta} (M_\odot\mathrm{yr}^{-1})]$  & $\log[\mathrm{SFR}_{\rm H\gamma} (M_\odot\mathrm{yr}^{-1})]$ & $\log[\mathrm{SFR}_{[\ion{O}{II}]} (M_\odot\mathrm{yr}^{-1})]$  \\  [0.4ex] 
  (1) & (11) & (12) & (13) & (14) & (15)  & (16) & (17)  \\  [0.4ex] 
  \hline 
96702  & - & $8.74_{-0.32}^{+0.22}$ & - & - & 1.08 $\pm$ 0.19 & - & 1.25 $\pm$ 0.47  \\  [0.6ex]
99389  & - & -                      & - & - & 1.91 $\pm$ 0.21 & - & 1.93 $\pm$ 0.45  \\  [0.6ex]
132869 & - & -                      & - & - & -               & - & -                \\  [0.6ex]
94458  & - & -                      & - & - & 1.68 $\pm$ 0.06 & - & 1.44 $\pm$ 0.44  \\  [0.6ex]
123801 & - & -                      & - & - & -               & - & 1.08 $\pm$ 0.47  \\  [0.6ex]
  \hline
  \end{tabular}

\bigskip \bigskip
\renewcommand\thetable{1} 
\caption{\textit{continued}}

\begin{tabular}{ccccccccccc}
  \hline \hline 
  objid & $\log[\mathrm{SFR}_{L_\mathrm{TIR}} (M_\odot\mathrm{yr}^{-1})]$  & $\log[\mathrm{SFR}_{\rm CIGALE} (M_\odot\mathrm{yr}^{-1})]$  & $\log[\mathrm{SFR}_{\rm CIGALE (10 Myrs)} (M_\odot\mathrm{yr}^{-1})]$  & $\log[\mathrm{SFR}_{\rm CIGALE (100 Myrs)} (M_\odot\mathrm{yr}^{-1})]$ \\  [0.4ex]
  (1) & (18)  & (19)  & (20)  & (21)   \\  [0.4ex]
  \hline 
96702  & 1.10 $\pm$ 0.12 & 0.87 $\pm$ 0.14 & 0.87 $\pm$ 0.14 & 0.83 $\pm$ 0.12 \\  [0.6ex]
99389  & 1.82 $\pm$ 0.05 & 1.59 $\pm$ 0.07 & 1.59 $\pm$ 0.07 & 1.58 $\pm$ 0.07 \\  [0.6ex]
132869 & 1.02 $\pm$ 0.03 & 0.74 $\pm$ 0.05 & 0.74 $\pm$ 0.05 & 0.73 $\pm$ 0.05 \\  [0.6ex]
94458  & 1.91 $\pm$ 0.02 & 1.61 $\pm$ 0.04 & 1.61 $\pm$ 0.04 & 1.59 $\pm$ 0.06 \\  [0.6ex]
123801 & 1.30 $\pm$ 0.06 & 0.95 $\pm$ 0.13 & 0.96 $\pm$ 0.13 & 0.94 $\pm$ 0.13 \\  [0.6ex]
  \hline
  \end{tabular}
\tablefoot{Column (1) is the unique identification number for each object in the Lockman--SpReSO catalogue. Column (2) is the spectroscopic redshift of the object. Column (3) is the stellar mass obtained from the SED fittings using CIGALE. Column (4) is the IR luminosity obtained from the SED fittings using CIGALE. Columns (5)--(8) are the colour excess calculated in this work (see Sect. \ref{sec:4}). Columns (9)--(13) are the metallicities derived in this work following different calibrations (see Sect. \ref{sec:5}). Columns (14)--(21) are the SFR derived in this work following different calibrations (see Sect. \ref{sec:6}).}

\end{sidewaystable*}


\section{Global relations} \label{sec:7}

\subsection{Main sequence in star-forming galaxies}

\begin{figure*}
    \includegraphics[trim= 1.5cm 1.5cm 1.5cm 0cm,width=\textwidth]{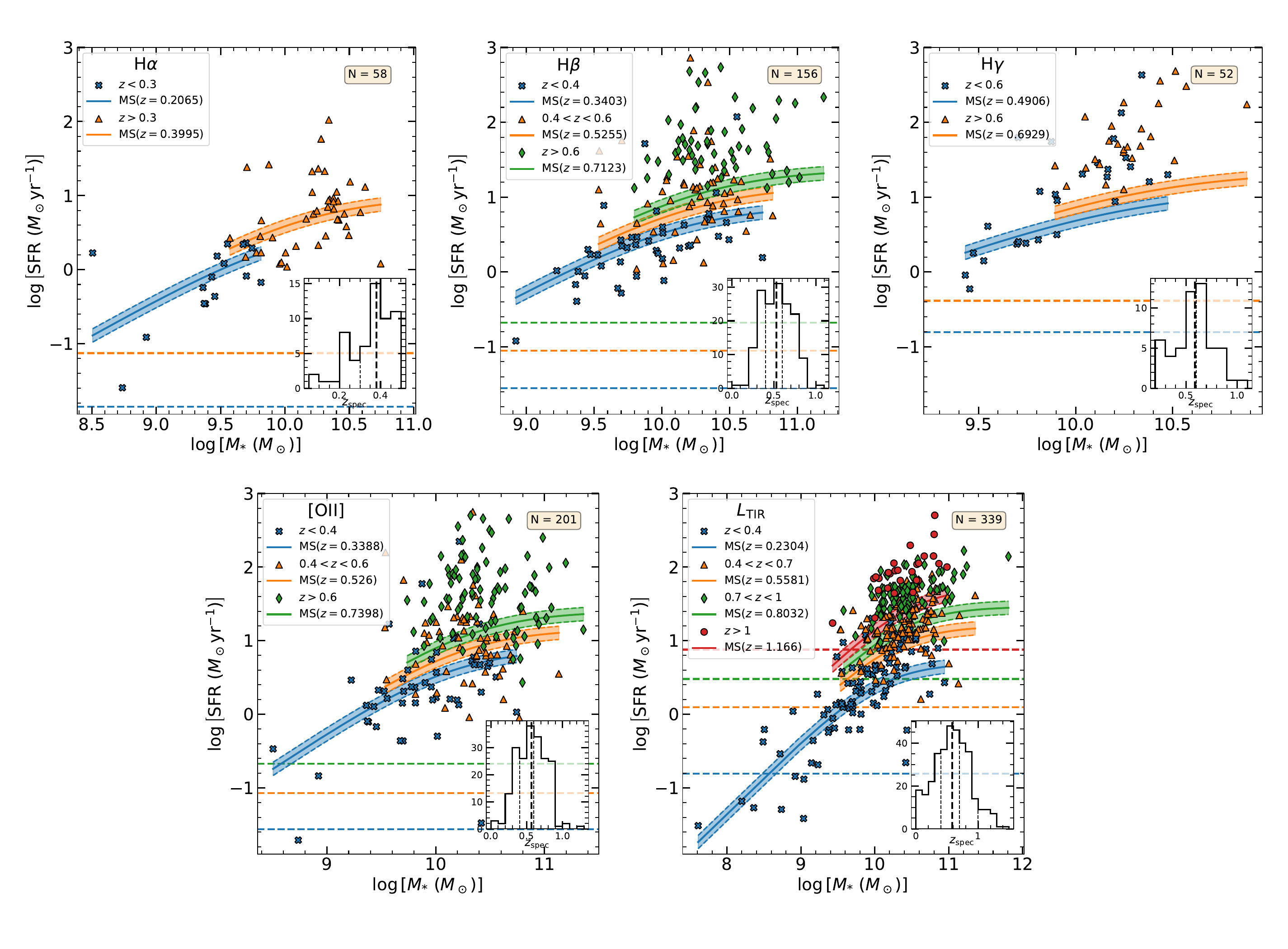}
   \caption{SFR--M relation for different SFR determination methods. From left to right and top to bottom, the SFR determined using H$\alpha$, H$\beta$, H$\gamma$, [\ion{O}{II}], and $L_\mathrm{TIR}$ fluxes is displayed. Inside each panel the sample was divided into redshift ranges. The solid lines represent the MS from \cite{Popesso23} evaluated at the mean redshift of each subsample, denoted by the points, using the same colour as the MS. The shaded areas and dashed lines represent 0.09 dex of scatter obtained by \cite{Popesso23}. The horizontal dashed lines designate the minimum detectable SFR for each redshift bin and SFR tracer, using the same colour scheme. The inset histograms show the redshift distribution for the complete sample in each panel. The vertical thick black dashed line indicates the redshift mean, and the vertical thin lines indicate the sample division boundaries. The numbers at the top indicate the total number of objects in each panel.}
    \label{fig:MS_zbin}
\end{figure*}

\begin{figure}
    \includegraphics[trim= 0cm 0cm 0cm 0cm,width=\linewidth]{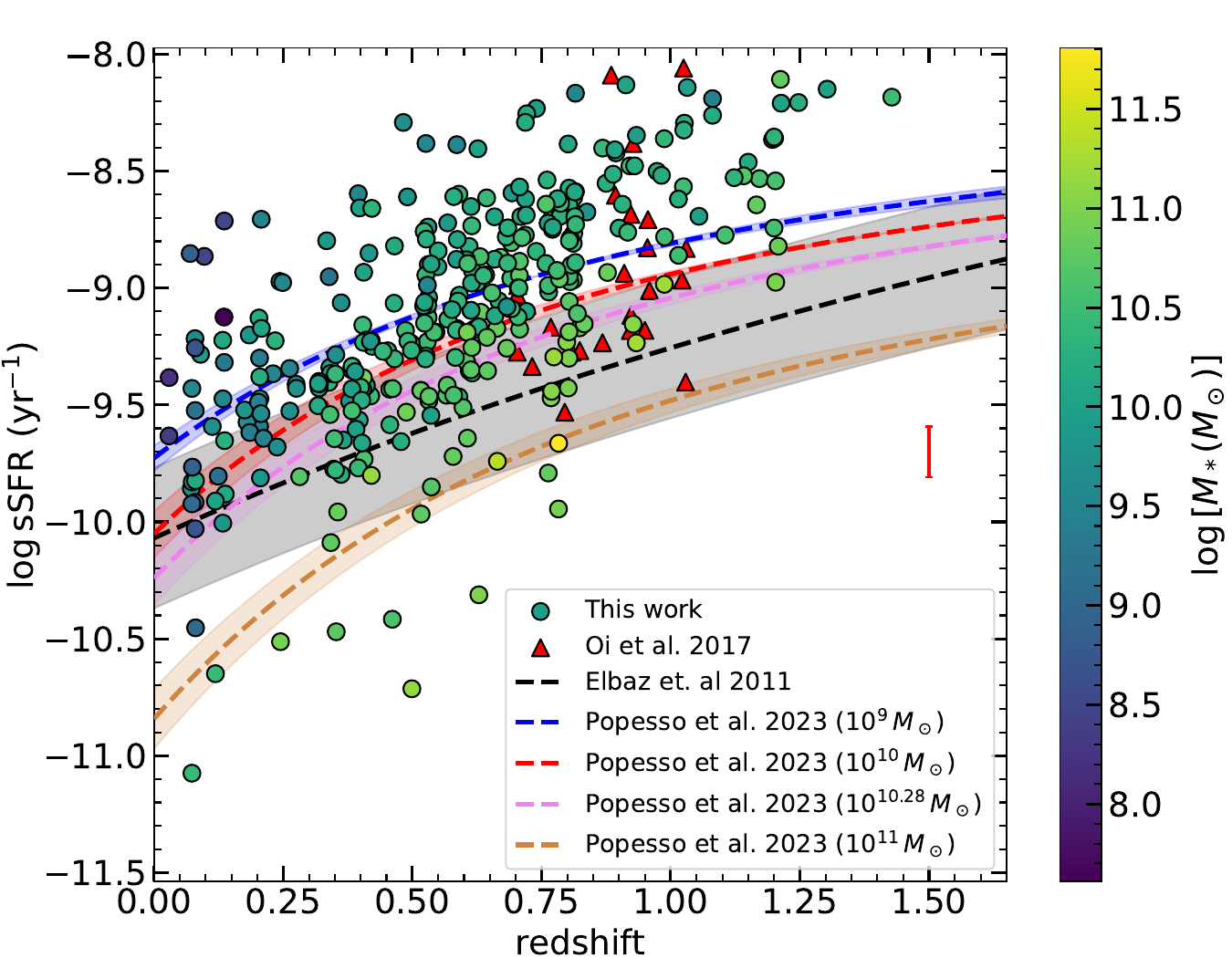}
   \caption{Redshift and $M_*$ evolution of the sSFR for SFG. The circles represent the Lockman--SpReSO galaxies, colour coded by $M_*$. The MS defined by \cite{Popesso23} is shown in magenta for the mean $M_*$ and in blue, red and brown for $\log M_* (M_\odot)$ set to 9, 10 and 11, respectively. The black line is the MS defined by \cite{Elbaz2011} for IR galaxies, assuming independence with $M_*$. The grey shaded region marks the area above which the galaxies are classified as starbursts. The Lockman sample tends to populate this region regardless of the MS used. The galaxies from \cite{Nagisa2017}, shown as red triangles, follow the behaviour of the Lockman--SpReSO objects. }
    \label{fig:MS_sSFR}
\end{figure}

\begin{table}
  \renewcommand\thetable{2} 
  \centering
  \caption{Mean values and errors of the shifts found with respect to the MS, obtained for the different SFR tracers. The mean values have been tabulated for objects with z=0.4 as separator, as well as the mean value for the whole sample.}
  \label{tab:shifts}
  \begin{tabular}{cccc}
  \hline \hline
     SFR   & \multicolumn{3}{c}{Mean shifts (dex)}  \\ \cline{2-4}
   tracer  &  $z<0.4$ & $z>0.4$ & All $z$ \\
  \hline
  H$\alpha$        & -0.05 $\pm$ 0.06  &  0.20 $\pm$ 0.10   &  0.04 $\pm$ 0.06    \\[0.25ex]
  H$\beta$         & -0.02 $\pm$ 0.06  &  0.49 $\pm$ 0.06   &  0.35 $\pm$ 0.05    \\[0.25ex]
  H$\gamma$        &  0.12 $\pm$ 0.12  &  0.65 $\pm$ 0.06   &  0.60 $\pm$ 0.06    \\[0.25ex]
  $[\ion{O}{II}]$  &  0.01 $\pm$ 0.08  &  0.38 $\pm$ 0.04   &  0.29 $\pm$ 0.04    \\[0.25ex]
  L$_\mathrm{TIR}$ &  0.10 $\pm$ 0.03  &  0.40 $\pm$ 0.02   &  0.32 $\pm$ 0.02    \\[0.25ex]
  \hline
  \end{tabular}
\end{table}

In this section, we examine the relationship between $M_*$ and the SFR of galaxies, as well as its possible evolution with redshift. There is a well-established positive correlation between $M_*$ and SFR, meaning that galaxies with higher stellar masses form stars at a higher rate than low stellar mass galaxies (see, for example, the comprehensive work of \citealt{Speagle2014} or more recently \citealt{Popesso23}, and references therein). Likewise, its evolution with redshift is widely acknowledged. For a given $M_*$ value, galaxies at higher redshifts form stars at a faster rate compared to galaxies at lower redshifts. This correlation is widely recognized as the main sequence of SFGs. The MS has been a subject of intense study, covering approximately five orders of magnitude in $M_*$ and spanning a redshift range from 0 to 6 (see table 4 of \citealt{Speagle2014} and table 1 of \citealt{Popesso23}).

In this paper, we adopt the MS model developed by \cite{Popesso23}, whose work involves a comprehensive synthesis of results from 28 studies focusing on the MS, aimed at investigating how this relation evolves over an extensive span of mass values, ranging from $10^{8.5}$ to $10^{11.5}$ $M_\odot$, and redshifts within the range $0 < z < 6$. The mathematical relationship they derived is as follows:
\begin{equation}\label{eq:MS}
    \log \mathrm{SFR} \left(M_*,t\right) = a_0 + a_1t -\log\left(1 + \left(  M_*/10^{a_2+a_3t}\right)^{-a_4}\right) 
\end{equation}
where $t$ is the cosmic time elapsed from the big bang in yr, $M_*$ is the stellar mass in solar masses, $a_0 = 2.693$, $a_1 = -0.186$, $a_2 = 10.85$, $a_3 = -0.0729$, and $a_4 = 0.99$.

In Fig.\ \ref{fig:MS_zbin}, we present the SFR derived through line fluxes and $L_{\rm TIR}$ plotted against $M_*$. As the MS described by \cite{Popesso23} depends on cosmic time, that is redshift,we have divided the sample into subsets by redshift ranges. For each of these subsets, we have overlaid the MS evaluated at the redshift of the subset. The MS is depicted with lines that match the colour of their respective subsets. The colour bands indicate the 0.09 dex scatter as determined by \cite{Popesso23}. For each subset, we have shown the minimum detectable SFR with a horizontal dashed line using the same colour code by redshift bin. The lowest SFR detectable using spectral line fluxes was obtained using the average EW for each line from \cite{Reddy2018}, in combination with the 1$\sigma$ level continuum limit $R_C <24.5$ mag set out in the Lockman--SpReSO project description \citep{Gonzalez2022}. For the calculation of the minimum detectable SFR based on $L_\mathrm{TIR}$, we set the limit values of 0.6 mJy and 2 mJy for the 100 $\mu$m and 160 $\mu$m bands from the \textit{Herschel}/PACS instrument, respectively, as described by the PEP team\footnote{\url{https://www.mpe.mpg.de/ir/Research/PEP/DR1/}} and converted to $L_\mathrm{TIR}$ using the \cite{Galametz2013} calibrations. Additionally, the inset histograms provide insights into the redshift distribution within each of the subsets. 

Irrespective of the method employed to derive the SFR, it is evident that the MS from \cite{Popesso23} fits well for objects at low redshifts ($z < 0.4$) where the mean shift showed by these objects from the MS is $\sim0.03$ dex. Table \ref{tab:shifts} displays the mean shift values and corresponding errors for each of the SFRs analysed in this study. The SFR derived from H$\gamma$ exhibits the largest distance. This is because there are not enough objects with $z<0.4$ to provide sufficient statistics for this SFR tracer. However, as we examine samples at higher redshifts, the trend shows that objects tend to be above the MS.
A significant fraction of the total sample (78\%) populates the starburst regime when the SFR derived from the $L_\mathrm{TIR}$ flux is examined, that is the region above the MS, and showing a mean shift from the MS of $\sim 0.4$ dex for objects at $z>0.4$. Table \ref{tab:shifts} shows the mean shift values for objects at $z>0.4$ based on the studied SFR tracers. The table also includes mean values for the full samples. The mean shift for H$\alpha$, which is mainly populated by low redshift objects, is very low. However, the other SFR determinations show a bigger median shifts from the MS as they compile objects at higher redshifts.

The objects in the sample with redshifts $z>0.4$ tend to be located near the starburst galaxy region. That is, they exhibit higher SFRs than that expected based on their $M_*$ and redshift, showing shifts of up to $\sim2$ dex, although \cite{Lee2017} found that galaxies departing from the MS by less than 0.6 dex may still be considered normal, not starburst. The SFRs obtained exceed the minimum detectable SFR by 1 to 2 orders of magnitude, proving that there is no noise-induced selection bias in our results. It is worth noting that the detection of galaxies heavily obscured by dust would be limited (up to 40\% of the initial sample), given that (Fig.\ \ref{fig:prop_distr}) most of our sample are LIRG or even FIR galaxies. Local ULIRGs are known to be outliers of the MS relation \citep{Elbaz2007}, as are starbursts \citep{Rodighiero2011, Guo2013}. For instance, it was found by \cite{Kilerci2014} that local ULIRGs with $M_*\sim10^{10.5-11.5}$ are more than an order of magnitude higher than the MS, and that the majority of ULIRGs are interacting pairs or post-mergers. 
However, the present study reveals that the intermediate redshift LIRGs exceed the MS, in average, by 0.5 dex, even at lower stellar masses than those examined by the authors. In contrast, the FIR and LIRGs at low redshift do not demonstrate this trait following the MS with a low scatter, as we have seen before. This indicates an evolutionary tendency of LIRGs, which has unique features compared to that of ULIRGs.

In the same way as the SFR--$M_*$ relation increases with redshift, the sSFR increases steadily up to $z\sim~2$ and then tends to flatten out \citep[][and references therein]{Speagle2014, Popesso23}. In Fig.\ \ref{fig:MS_sSFR} we have plotted the sSFR, using the SFR derived by $L_\mathrm{TIR}$, against redshift and coloured by $M_*$ for the Lockman--SpReSO and \cite{Nagisa2017} objects. The galaxies from \cite{Nagisa2017} are a sample of AKARI-detected mid-IR SFG at $z \sim 0.88$ with Subaru/FMOS spectroscopic observations. The MS defined by \cite{Popesso23} is also plotted with  $\log M_* (M_\odot)$ set to the median value of the sample (10.28), and 9, 10 and 11, showing the evolution with redshift and $M_*$. The relationship of \cite{Elbaz2011} shown in Fig.\ \ref{fig:MS_sSFR} was produced by examining infrared SEDs for a sample of objects in the redshift range $0<z<2.5$. For the study, they assumed a slope of 1 in the SFR--$M_*$ relation and constant over the whole redshift range; that is, independent of mass:
\begin{equation}
    \mathrm{sSFR}\,(\mathrm{yr}^{-1}) = 26\cdot10^{-9} \times t^{-2.2}
\end{equation}
where $t$ is the cosmic time. They mark as starbursts those galaxies which, for the same redshift, have an sSFR twice that of the MS, represented by the upper edge of the grey-shaded area. As for the case of the SFR--$M_*$ relation, the Lockman--SpReSO objects have sSFRs that tend to be higher than the MS shown in the figure, so we are studying galaxies with very intense outbursts of star formation; that is, starburst galaxies. However, although the data follow the general trend of decreasing sSFR with increasing $M_*$, no evident flattening can be observed with redshift, and only the most massive galaxies seem to fit the \cite{Popesso23} and \cite{Elbaz2011} relations, while for galaxies with stellar mass values that are below the median, sSFR is higher than the models showing a median scatter of 0.31 dex and 0.56 dex from \cite{Popesso23} and \cite{Elbaz2011}, respectively.

\subsection{Mass--metallicity relation for infrared galaxies}

\begin{figure*}
    \includegraphics[trim= 0cm 0cm 4.5cm 0cm,width=\textwidth]{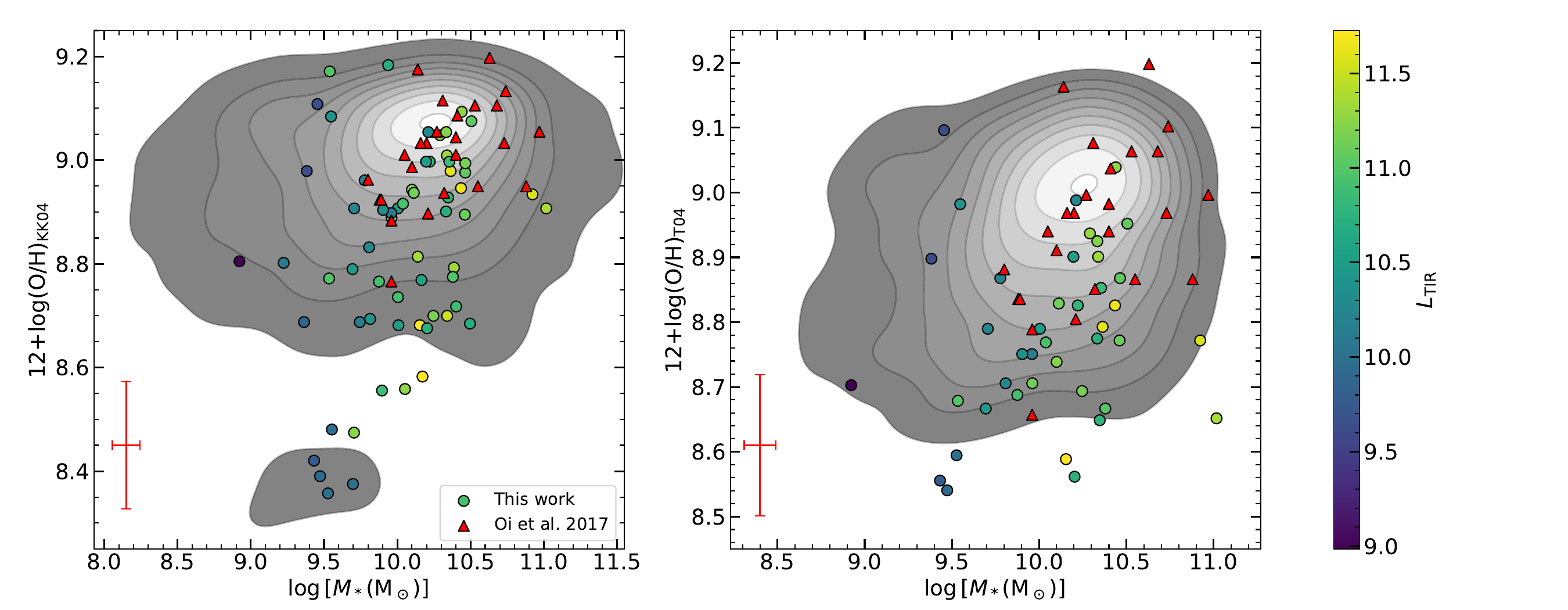}
   \caption{MZR using the metallicity parameterisation from \citet[left]{Kobulnicky2004} and \citet[right]{Tremonti2004}. In both panels the galaxies from Lockman--SpReSO are colour coded according  to $L_\mathrm{TIR}$. The red triangles are the galaxies from the paper by \cite{Nagisa2017}. The shadows and contours in the background represent the merged galaxies from the OSSY catalogue \citep{OSSY2011} and the HELP catalogue \citep{Shirley2019}. The mean error is shown in red at the bottom left.}
    \label{fig:MZR_plot}
\end{figure*}

\begin{figure*}
    \includegraphics[trim= 0cm 0cm 4.5cm 0cm,width=\textwidth]{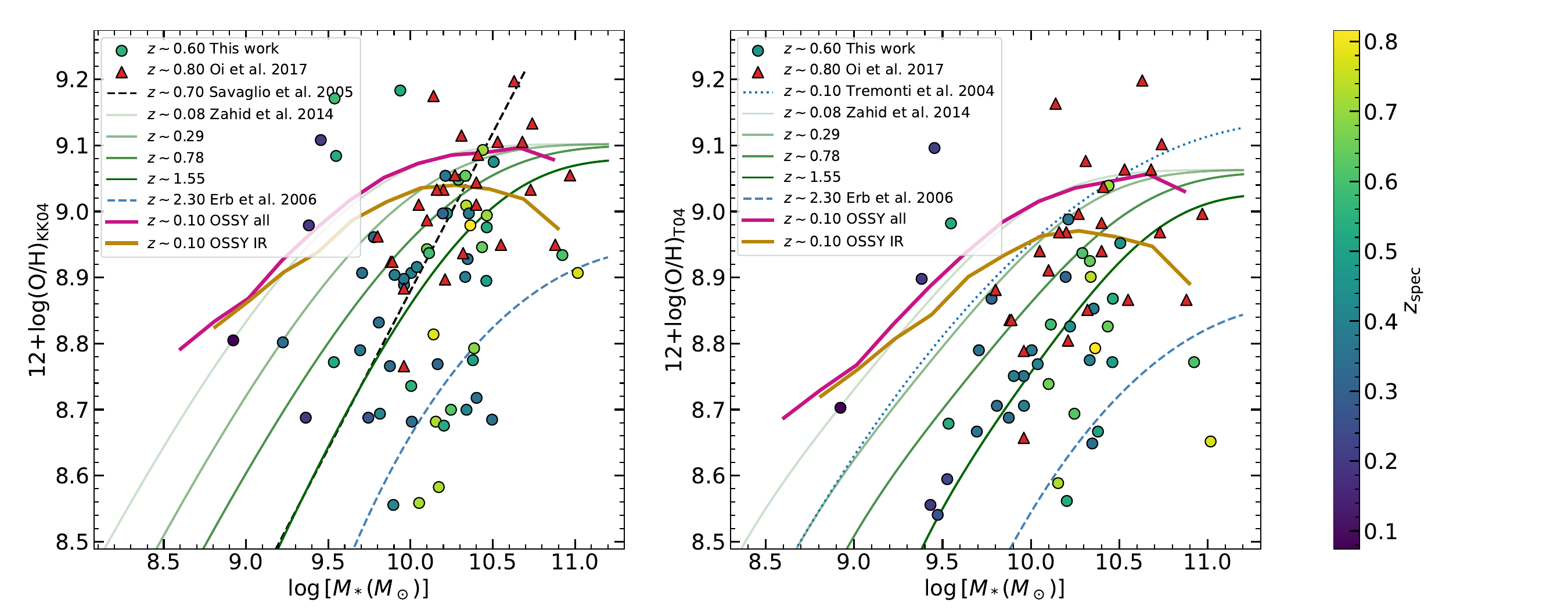}
   \caption{MZR diagram using the \citet[left]{Kobulnicky2004} and \citet[right]{Tremonti2004} metallicity calibrations. The blue dots represent our Lockman--SpReSO data at a median redshift $\sim0.6$ and the red triangles represent the \cite{Nagisa2017} IR galaxies at a median redshift $\sim0.8$. The magenta line represents the median metallicity for each mass binning of the whole OSSY catalogue, while the brown line represents the same binning but only for the OSSY objects with IR information in HELP. The green lines represent the MZR model from \cite{Zahid2014} applied to galaxies from the SDSS ($z\sim0.08$), SHELS ($z\sim0.29$), DEEP2 ($z\sim0.78$) and COSMOS ($z\sim1.55$) surveys. The black dashed line represents the linear MZR model obtained by \cite{Savaglio2005} for 60 galaxies with an average redshift of $z\sim0.7$. In both the OSSY catalogue and the Lockman objects, it can be seen that, except at the higher masses of the sample, IR objects tend to be less metallic than optical objects at the same redshift. The blue dashed line shows the MZR obtained by \cite{Erb2006} for a sample of UV-selected SFGs at $z\sim2.3$.}
    \label{fig:MZR_comparison}
\end{figure*}

The existence of a relation between $M_*$ and metallicity of galaxies is well known in a wide range of spectroscopic redshifts ($0<z<3$), with lower-mass galaxies having lower metallicity and more massive galaxies having higher metallicity. However, the evolution of this relation with redshift is still a matter of debate. This discussion also extends to the analysis of IR objects and the evolution of the $M_*$--metallicity relation for these objects and the differences with optically selected ones.

As a benchmark to low redshift ($z\sim0.1$), we have used the OSSY catalogue \citep{OSSY2011}, an improved and quality-assessed set of emission and absorption line measurements in SDSS galaxies. The $M_*$ of these objects were taken from SDSS Data Release 10 \citep{Ahn2014}, obtained by SED fits using the Sloan photometric bands\footnote{The IMF used by \cite{Ahn2014} in the SED fitting process is that of \cite{Kroupa2001}. To transform from \cite{Kroupa2001} to \cite{Chabrier2003} a multiplicative factor of 0.94 or -0.02 in dex must be applied.}. In addition, to make the comparison between IR-selected samples of galaxies, we matched the OSSY catalogue with the \textit{Herschel} Extra-galactic Legacy Project (HELP) database \citep{Shirley2019}. The HELP group has merged the information from the various fields observed by \textit{Herschel} to create a general catalogue in which, in addition to the IR information of the objects, they have added all the information in other photometric bands obtained by other surveys, from UV to FIR. Using this information, they performed a SED fit analysis of the sources catalogued in the HELP database using CIGALE software, providing new measurements of $M_*$, $L_{\rm TIR}$ and extinction, among other parameters \citep{Malek2018}.

To analyse the MZR, we used the metallicity obtained by the iterative method of \citetalias{Kobulnicky2004} and \citetalias{Tremonti2004}, both based on the R$_{23}$ tracer, since it provides the most complete set in terms of number of objects and is widely used in the literature, together with the $M_*$ obtained by the SED fitting process in \cite{Gonzalez2022}. Figure \ref{fig:MZR_plot} shows the MZR diagram obtained for both metallicity estimators. The metallicity of \cite{Nagisa2017} was calculated using the N2 tracer calibrated by \citetalias{Pettini2004}. For comparison with the Lockman--SpReSO sample, the metallicities have been converted from N2 to R$_{23}$ by performing a calibration using the full OSSY sample and a second-order polynomial fit, see Appendix \ref{sec:appendixB} for details. It can be seen, from both panels, that the metallicities derived for the Lockman--SpReSO and \cite{Nagisa2017} samples are compatible, and tend to be lower than those of the OSSY sample, with the exception being the highest $M_*$, which corresponds to the area with the densest concentration of OSSY galaxies. Moreover, this result agrees with that obtained by \cite{Nagisa2017}, who find that the metallicities for their IR-selected objects are compatible with those obtained by \citetalias{Tremonti2004} for normal SFGs in the local Universe ($z\sim0.1$) and higher than those of \cite{Zahid2011} for a sample of DEEP2 galaxies with $z\sim0.78$. However, the \cite{Nagisa2017} data represent the highest redshift and the highest stellar mass of the Lockman--SpReSO sample. At lower  stellar masses the metallicity is lower by $\sim 0.25\, \mathrm{dex}$. Both properties show a positive Spearman's rank correlation coefficient of 0.36 at a significance of 3.5$\sigma$ for the metallicity of \citetalias{Kobulnicky2004} and 0.34 at a significance of 2.9$\sigma$ for the metallicity of \citetalias{Tremonti2004}. In addition, we proceeded to test whether IR objects behave differently from optical objects when studying the MZR. To this end, we analysed the behaviour of the IR properties in MZRs obtained from samples of objects not selected according to their IR flux.

In Fig.\ \ref{fig:MZR_comparison} we show the MZR diagram using the metallicities from \citetalias{Kobulnicky2004} upper branch, and from \citetalias{Tremonti2004} for the Lockman--SpReSO and \cite{Nagisa2017} objects. In addition, we have binned the $M_*$ for the OSSY catalogue, for both the full catalogue and the catalogue merged with HELP, considering the median metallicity in each bin (only bins those with more than 40 galaxies are represented). This reveals that the metallicity of the bins with an IR detection is lower than that obtained for the full catalogue in the two metallicity tracers studied, as can be seen in the two panels of Fig.\ \ref{fig:MZR_comparison}. For a sample of nearby LIRGs and ULIRGs at a redshift of around 0.1, \cite{Rupke2008} also found that these types of galaxies exhibited a median metallicity offset of 0.4 dex from the MZR discovered \citetalias{Tremonti2004}.

The \cite{Zahid2014} model, originally computed using an IMF of \cite{Chabrier2003} and the metallicity calibration of \citetalias{Kobulnicky2004}, is also included in Fig.\ \ref{fig:MZR_comparison} for local SDSS galaxies ($z\sim0.08$) and distant galaxies from the SHELS ($z\sim0.29$), DEEP2 ($z\sim0.78$) and COSMOS ($z\sim1.55$) surveys. The linear relationship obtained by \cite{Savaglio2005} for the MZR is also shown in Figure \ref{fig:MZR_comparison}. They studied 60 SFGs in the redshift range $0.4<z<1.0$ using the metallicity calibration of \citetalias{Kobulnicky2004} and an IMF \citet[1.13 times that of \citealt{Chabrier2003}]{Baldry2003}. As a benchmark for high redshift we have shown the MZR from the \cite{Erb2006} paper, an analysis of 87 rest-frame UV-selected SFG with a mean redshift $z\sim2.23$ from Keck/LRIS observations that also confirms the evolution of the MZR with cosmic time. The \citetalias{Kobulnicky2004} to \citetalias{Tremonti2004} metallicity transformation was calibrated in the same way as was done previously for the metallicity from \cite{Nagisa2017}, as shown in Appendix \ref{sec:appendixB}. It can be seen that the result obtained by \cite{Zahid2014} for their SDSS sample and our result for the full OSSY sample are in good agreement, while again the OSSY IR objects have lower metallicities than the full OSSY sample. 
The MZR model at $z\sim 1.55$ is the best fit for the lower-mass Lockman--SpReSO data ($M_*\lesssim 10^{10.2}\, M_\odot$) showing a median dispersion from the MZR of 0.04 dex, while for \cite{Nagisa2017} and the Lockman--SpReSO data of similar masses both models $z\sim 1.55$ and $z\sim 0.78$ fit the behaviour of the data well, showing the same absolute dispersion from both MZRs (0.07 dex). This is due to the short evolution of the MZR and the dispersion of the data in this range of $M_*$.
This again indicates that the metallicities of Lockman--SpReSO data at $M_*\lesssim 10^{10.2}\, M_\odot$ are lower than predicted by the models, especially when we analyse the MZR using the \citetalias{Tremonti2004} metallicity calibration, where the Lockman--SpReSO objects show a median dispersion of 0.02 dex with respect to the MZR at $z\sim1.55$. At higher $M_*$, the redshift evolution of the MZR is smaller. It is evident in the displayed MZRs in Fig.\ \ref{fig:MZR_comparison} that the most massive galaxies attain almost their current metallicity at redshift $z\sim1$ \citep{Zahid2014,Maiolino2019}.

\subsection{The $M_*$--SFR--metallicity relation}
\begin{figure}
    \includegraphics[trim= 0cm 1.5cm 0cm 0cm,width=\linewidth]{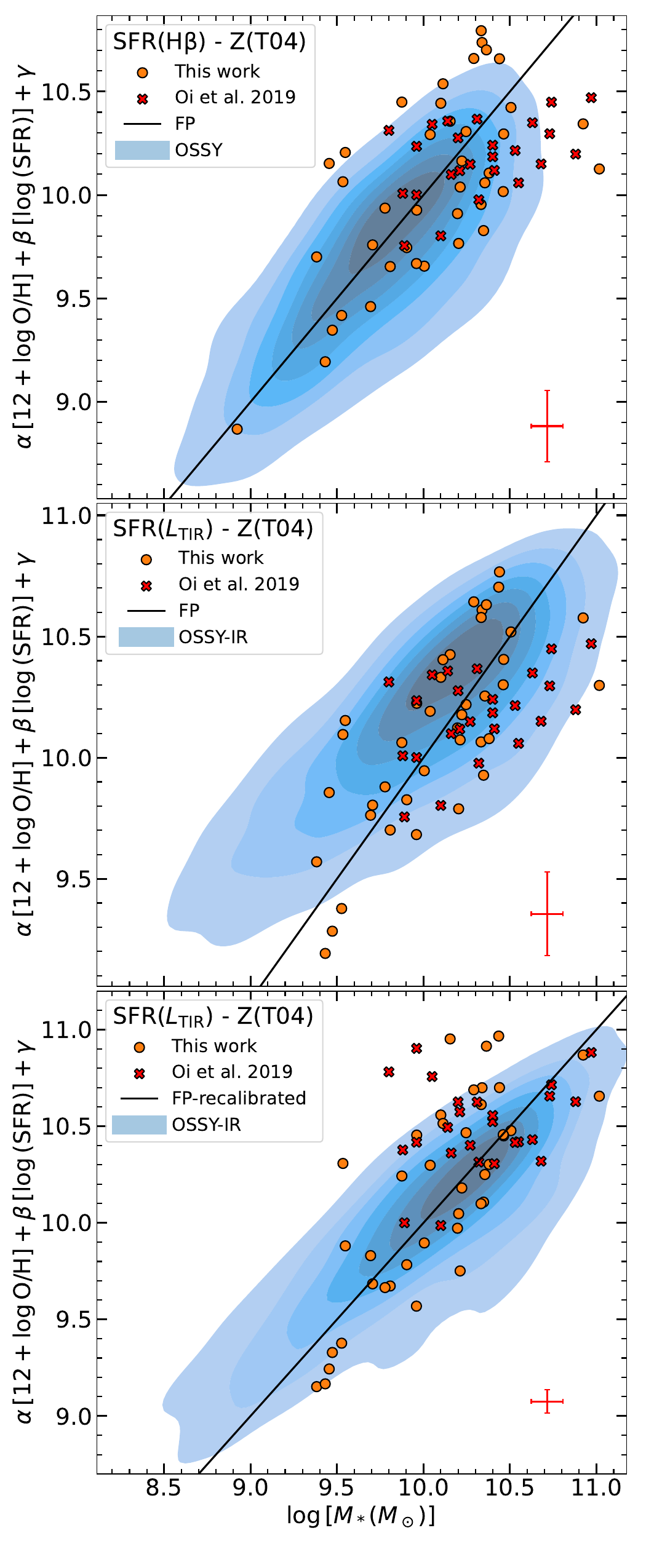}
   \caption{Projection of the FP onto the $M_*$ coordinate against the observed $M_*$. In the top panel we have used the SFR derived from the H$\beta$ flux, the metallicity using the \cite{Tremonti2004} calibration and the \cite{Lara2013} parameterisation. The orange dots are the Lockman--SpReSO SFGs, the red crosses are the \cite{Nagisa2017} data and the blue contours at the bottom are the OSSY data. The middle panel shows the same, but with the IR-derived SFR. In the lower panel we show the recalibration of the FP using the OSSY data with IR information to obtain a calibrated FP with the $L_{\rm TIR}$-derived SFR. A more detailed discussion is given in the text.}
    \label{fig:FMR}
\end{figure}

As we have seen above, there is a strong relationship between $M_*$ and SFR, as well as between $M_*$ and metallicity, both of which have been extensively studied in the literature. However, it is only relatively recently that a clear dependence between SFR and metallicity has become evident. The initial hints of this relationship were discovered by \cite{Ellison2008}, who found a slight connection between sSFR and metallicity. Shortly afterwards, almost simultaneously and independently, \cite{Lara2010} and \cite{Mannucci2010} found and described the mutual relationship between SFR, $M_*$, and metallicity.

\cite{Lara2010} used a complete magnitude-limited sample of SFGs, with the $r$-band falling in the range 14.5 to 17.77 mag, sourced from the SDSS-DR7 catalogue. This sample spanned a redshift range of $0.04 < z < 0.1$.  Using the $M_*$ as the dependent variable on SFR and metallicity, they fitted a plane to the distribution formed in the 3D space of these parameters. With a scatter of 0.16 dex in their fit, they found the existence of a clear relationship between the three galaxy properties, which they named the Fundamental Plane (FP). Furthermore, when \cite{Lara2010} compared their findings with data from studies at higher redshifts, extending up to $z\sim3$, they ascertained that the FP exhibited no evolution with redshift. In subsequent papers \cite{Lara2013, Lara13b}, revisited the FP. This revision involved expanding the studied sample by adding data from the GAMA survey, which is two orders of magnitude deeper than the SDSS (up to $z\sim0.35$). They also applied principal component analysis (PCA) in their investigation, resulting in a reduction in the scatter within the Fundamental Plane. The FP offers the ability to estimate the $M_*$ of SFGs using both SFR and metallicities, with a dispersion of 0.2 dex, as demonstrated by \cite{Lara2013}.

Using the SDSS--DR7 sample for the redshift range $0.07<z<0.3$, \cite{Mannucci2010}, and the \cite{Mannucci2011} application for a low-mass sample, they determined that SFGs delineate a surface in 3D space defined by $M_*$, metallicity, and SFR. This relation was designated the Fundamental Metallicity Relation (FMR). The scatter they identified was $\sim0.05$ dex, a value in line with the uncertainties inherent in the galaxy properties analysed. \cite{Mannucci2010} used metallicity as the dependent variable in relation to $M_*$ and SFR.

The findings from both the FP and the FMR point to a robust connection for SFGs among SFR, $M_*$, and metallicity. Equally significant is the observation that this connection remains unaltered with redshift, persisting up to $z\sim3$, which is also confirmed by several independent works \citep[see for example][among many others]{Hunt2012,Cresci2019,Sanders2021}.

In this paper, to study the relationship between SFR, $M_*$ and metallicity, we used the plane described by \cite{Lara2010}, where $M_*$ is studied as a function of SFR and metallicity using the following expression:
\begin{equation} \label{eq:FP}
    \log M_* = \alpha\,[\meta] +\beta\,[\log \mathrm{SFR}] + \gamma
\end{equation}
where $\alpha=1.3764$, $\beta=0.6073$, and $\gamma=-2.5499$ obtained from the revision of the FP by \cite{Lara2013}.

In the top panel of Fig.\ \ref{fig:FMR} we have plotted the fundamental plane (Eq.\ \ref{eq:FP}) for the metallicity calibration obtained by \citetalias{Tremonti2004} and SFR obtained using the H$\beta$ luminosity. It can be seen that the fundamental plane reproduces very well both the sample of local OSSY objects, represented by the background contours, and the sample of IR-selected Lockman--SpReSO galaxies when using the SFR derived from the H$\beta$ flux, obtaining an average scatter of the plane of 0.20 dex, which is only slightly larger than the average uncertainty found for the $M_*$ determined by SED fits (0.1 dex). However, when using $L_{\rm TIR}$, the Fundamental Plane does not reproduce the behaviour of the data, as can be seen in the middle panel of Figure \ref{fig:FMR}. This is due to the definition of the FP by \cite{Lara2013}, which specifically uses the H$\alpha$ flux and the \citetalias{Tremonti2004} metallicity for its construction. In order to carry out a fair comparison, it is necessary to recalibrate Eq.\ \ref{eq:FP}, because we have to take into account that when analysing $L_{\rm TIR}$ and the Balmer lines, we are studying not only different star formation timescales, but also different regions of the galaxies, although they are related. Assuming that the metallicity derived using optical lines is representative of the whole galaxy, and using the OSSY sample with IR information, we have refitted Eq.\ \ref{eq:FP}. We have used the SFR derived from the IR luminosity and $M_*$ obtained by the HELP team using SED fits with CIGALE \citep{Malek2018,Shirley2019}. The new parameters obtained are $\alpha=0.3640$, $\beta=0.9071$ and $\gamma=6.1029$. The result is shown in the lower panel of Fig.\ \ref{fig:FMR}, where we have plotted the recalibration of the Fundamental Plane with the IR derived SFR and \citetalias{Tremonti2004} metallicity calibrator. It can be seen that the FP now reproduces very well the behaviour of the Lockman--SpReSO data with an average scatter of 0.17, although for the \cite{Nagisa2017} data there seems to be a wider scatter, mainly owing to the metallicity transformation used for comparison. This result shows that the Fundamental Plane makes it possible to calculate the $M_*$ of galaxies with a low level of uncertainty.

\cite{Salim2014} re-analysed the SDSS galaxies and found that for $M_*\gtrsim10.5\,M_\odot$ the relationship between sSFR and metallicity appears to be weak or non-existent, a result also found in simulations by \cite{Matthee2018}, who argue that it is due to contamination by AGN. Although in this paper the FP study using the SFR from H$\beta$ flux (top panel in Fig. \ref{fig:FMR}) seems to support this trend, the FP study using the SFR from $L_\mathrm{TIR}$ flux (bottom panel in Fig. \ref{fig:FMR}) does not support this trend; and neither does our sample include AGN.

From the previous two subsections, we have shown that Lockman--SpReSO SFGs tend to have lower metallicities than normal galaxies and SFRs above the MS for redshifts $z>0.4$, typical of starburst galaxies. Despite these unusual properties, the Fundamental Plane remains valid, although it was formulated using samples of objects very different from those studied here. Moreover, neither do we find any difference in the observed trends between the different SFR tracers. This also supports the non-evolutionary theory of the Fundamental Plane, since no redshift trends are evident. This result is in agreement with that found by \cite{Hunt2012} for a sample of $\sim1000$ extreme and rare objects. They selected local quiescent SFGs and blue compact dwarfs, luminous compact emission line galaxies at $z=0.3$ and Lyman-break galaxies spanning a redshift range of $1<z<3$. In plots such as MZR or MS, these objects appear as outliers owing to their extreme properties. However, these objects follow the FP with good accuracy, a result that extrapolates the relationship between SFR, $M_*$ and metallicity to extreme class objects.

The most accepted explanation for the non-evolution of the FP lies in the balance between SFR and metallicity at different stages of galaxy evolution \citep[see the review by][and references therein]{Maiolino2019}. In general, galaxies at high redshift are composed of stars formed from poorly processed gas (that is of low metallicity), which is also associated with high SFR, as observed at these evolutionary stages. On the other hand, if we look at galaxies in the local Universe, stars are formed from highly processed material (that is, of high metallicity). Moreover, the availability of gas to form stars in local galaxies is more limited than in high-redshift galaxies, which implies lower SFR. This shows the balance, with high redshift galaxies having high SFRs at low metallicities, and local galaxies having low SFRs at higher metallicities. The fact that the evolution of SFR and metallicity with redshift go in opposite directions, helps to explain the non-evolution of the Fundamental Plane, although the parameterization can vary depending on the timescale of the SFR indicator used, as can be seen in the bottom panel of Figure \ref{fig:FMR}.


\section{Summary and conclusions}\label{sec:8}

In this paper, we present the first study of the SFGs of the Lockman--SpReSO project. This project is an optical spectroscopic follow-up of \textit{Herschel} FIR-selected sources with optical counterparts of $R_C<24.5$ mag. The scope of the present work was to determine fundamental parameters such as extinction, metallicity, and SFR, and to study the relationships among them. To this aim, we used the FIR-selected galaxies for which the spectroscopic redshift was determined by \cite{Gonzalez2022}, with the result that, in a total of 409 objects, the redshifts lay in the range $0.03<z<4.96$. The objects in the Lockman--SpReSO catalogue have photometric information over a wide spectral range, from X-ray to FIR bands. Apart from the optical spectroscopy, the derived redshifts and line fluxes, SED fits were also performed, from which, among other properties, $M_*$ and $L_\mathrm{TIR}$ of the galaxies were obtained. These results have yielded a spectroscopic study with a higher number of carefully FIR-selected SFGs.

For the study, 69 objects (17\%) identified as AGN based on the criteria outlined in Sec. \ref{sec:3} were excluded from the sample. This percentage is notably higher than that of local samples but is dependent largely on the selection criteria used. Finally, the resulting sample consists of 340 SFGs, with almost half of the sample being LIRGs.

From the analysis of the relationship between $M_*$ and SFR, it resulted that SFGs at low redshifts, $z<0.4$, follow the MS defined by \cite{Popesso23}, with a shift of $\sim0.1$ dex, based on the emission line fluxes and $L_{\rm TIR}$ SFRs. However, at higher redshifts, the sample presents significant evolution by increasing the fraction of starburst galaxies, with 78\% of the galaxies falling into this classification using the SFR derived from $L_{\rm TIR}$. The shift from the MS is approximately 0.4 dex for each object. Therefore, SFGs from Lockman--SpReSO exhibiting significant burst of star formation, as would be expected for a FIR-selected sample. This result can also be observed when analysing the sSFR and its evolution over cosmic time, even when compared to the MS, such as that of \cite{Elbaz2011}, which is designed for IR objects, although they assume that the MS is independent of $M_*$. However, in the present sample, no apparent flattening of sSFR with redshift for $\log M_* (M_\odot)\gtrsim10.5$ is observed.

The MZR relationship, derived using \citetalias{Tremonti2004} and \citetalias{Kobulnicky2004} metallicity calibrators, compared with data from \cite{Nagisa2017}, shows that both samples are compatible. The comparison with the OSSY sample \citep{OSSY2011} containing IR information in the HELP database \citep{Shirley2019} shows that Lockman--SpReSO galaxies have lower metallicities. Moreover, comparing with MZR from the literature \cite{Tremonti2004,Savaglio2005,Erb2006,Zahid2014}, Lockman--SpReSO FIR-selected SFGs exhibit lower metallicities than anticipated for their redshift and $M_*$, since the MZR defined by \cite{Zahid2014} at $z\sim1.55$ fits Lockman--SpReSO galaxies with a scatter of 0.1 dex. This result is particularly evident for masses  $M_*\lesssim10^{10.2}\,M_\odot$. At higher masses, where the Lockman--SpReSO and \cite{Nagisa2017} data overlap, due to the dispersion of the data and the short range of evolution of the MZRs at those $M_*$, the metallicity is equally fit by the MZR at $z\sim1.55$ and at $z\sim0.78$, closer to the mean redshift of the sample. Nonetheless, there is a limited number of Lockman--SpReSO galaxies in this region, while the majority of \cite{Nagisa2017} galaxies are present in this area. The present study also conducted the MZR comparison using both OSSY galaxies with IR photometric information, as well as the entire OSSY sample, showing that local galaxies with IR information also showcased decreased metallicities. This finding confirms the notion that IR galaxies have a tendency to show lower metallicities than optical galaxies.

Finally, by incorporating the SFR as an additional parameter in the MZR, which then becomes the FP, the dispersion is substantially reduced. The 3D correlation has been investigated following the work of \cite{Lara2010}, who established the FP, where $M_*$ is the  dependent variable on both SFR and metallicity. Based on our analysis of the \citetalias{Tremonti2004} metallicity and the SFR derived from H$\beta$ flux, we have concluded that the FP is valid for the Lockman--SpReSO LIRG sample, despite the objects exhibiting strong SFR outbursts and lower metallicities compared to optical galaxies, showing a median scatter about the FP of $\sim0.20$ dex. However, a recalibration of the FP is required to be able to use the SFR derived from $L_\mathrm{TIR}$. It has been established, then, that the FP is valid for LIRG objects. Nevertheless, when using $L_\mathrm{TIR}$ the known saturation of the relation $\log M_*(M_\odot)\gtrsim10.5$ is not observed, contrary to the findings of \cite{Salim2014} and \cite{Matthee2018}, for non-IR-selected galaxies. This could point towards an evolution of the more massive fraction of the sample, in the sense of decreasing present-day star formation with respect to the averaged star formation in the past. The balance between the SFR excess and the metallicity deficit justifies the applicability of the FP \citep[among many others]{Lara2010,Lara13b,Maiolino2019}, further supporting the reasoning behind its non-evolution.

\begin{acknowledgements}

We thank the anonymous referee for their useful report. 
This work was supported by the Evolution of Galaxies project, of references: 
PRE2018-086047,
AYA2017-88007-C3-1-P, 
AYA2017-88007-C3-2-P, 
AYA2018-RTI-096188-BI00, 
PID2019-107408GB-C41, 
PID2019-106027GB-C41 and, 
PID2021-122544NB-C41,
PID2022-136598NB-C33,
within the \textit{Programa estatal de fomento de la investigación científica y técnica de excelencia del Plan Estatal de Investigación Científica y Técnica y de Innovación (2013-2016)} of the Spanish Ministry of Science and Innovation/State Agency of Research MCIN/AEI/ 10.13039/501100011033 and by `ERDF A way of making Europe'. 
This article is based on observations made with the Gran Telescopio Canarias (GTC) at Roque de los Muchachos Observatory on the island of La Palma, with the Willian Herschel Telescope (WHT) at Roque de los Muchachos Observatory on the island of La Palma and on observations at Kitt Peak National Observatory, NSF's National Optical-Infrared Astronomy Research Laboratory (NOIRLab Prop. ID: 2018A-0056; PI: Gonz\'alez-Serrano, J.I.), which is operated by the Association of Universities for Research in Astronomy (AURA) under a cooperative agreement with the National Science Foundation. This research has made use of the NASA/IPAC Extragalactic Database (NED), which is funded by the National Aeronautics and Space Administration and operated by the California Institute of Technology.
J.N acknowledge the support of the National Science Centre, Poland through the SONATA BIS grant 2018/30/E/ST9/00208.
YK aknowledges support from PAPIIT grant 102023.
ICG and EB acknowledge financial support from DGAPA-UNAM grant IN-119123 and CONAHCYT grant CF-2023-G-100.
The authors thank Terry Mahoney (at the IAC's Scientific Editorial Service) for his substantial improvements of the manuscript.

\end{acknowledgements}

\bibliographystyle{aa}
\bibliography{biblio} 

\newpage
\onecolumn
\begin{appendix} 
\section{Branch selection criteria} \label{sec:appendixA}
\begin{figure*}[h!]
    \includegraphics[trim= 0cm 0cm 0cm 0cm,width=\textwidth]{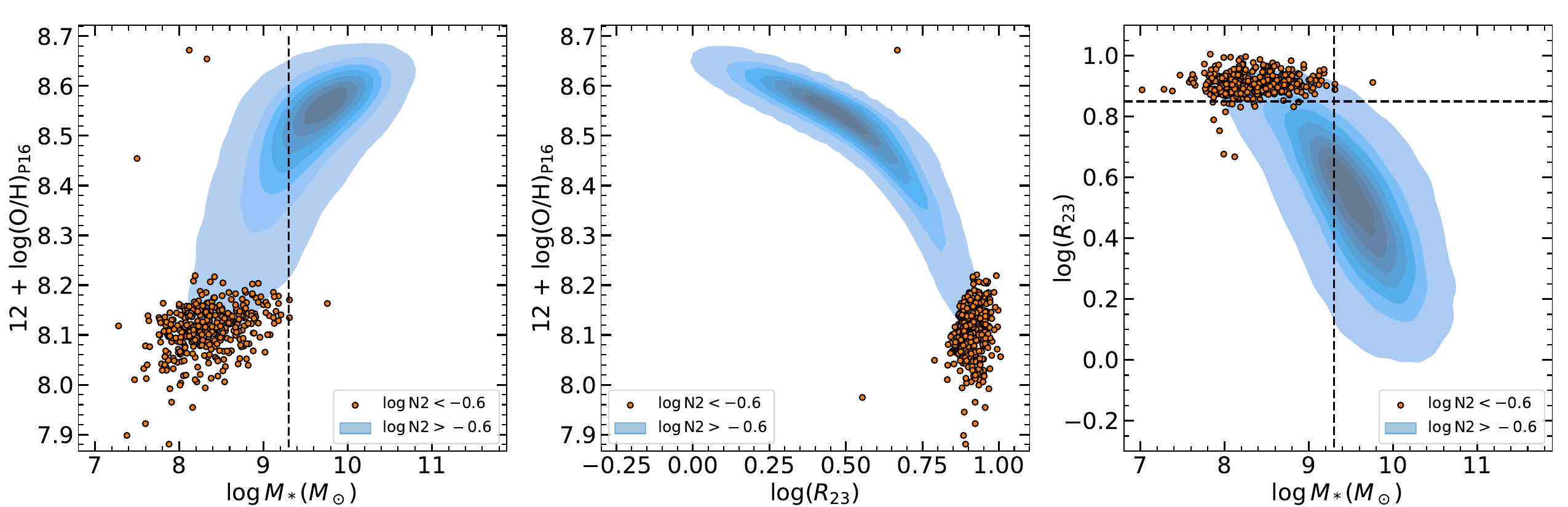}
   \caption{Criteria for separating galaxies into the upper and lower branches. On the left we have plotted the metallicity obtained with the \cite{Pilyugin2016} calibration against $M_*$. In orange we have plotted the OSSY data that meet the $\log N_2<-0.6$ criterion, defined by \cite{Pilyugin2016} as the branch separation. The blue contours in the background represent the OSSY objects in the upper branch, and the vertical dashed line marks a $M_*$ value of $\log M_*(M_\odot)=9.3$, which seems to mark a limit for the objects in the lower branch. In the middle panel we show the relationship between the same metallicity and the parameter $R_{23}$. In the right panel we have plotted the parameter $R_{23}$ against $M_*$. The vertical line at $\log M_*(M_\odot)=9.3$ and the horizontal line at $R_{23}=0.85$ mark the region where the lower branch objects tend to be.}
    \label{fig:branch_selection}
\end{figure*}

\citetalias{Pilyugin2016} use the value of the $N_2$ parameter $\left([\ion{N}{II}]\, \lambda\lambda6548.84/\mathrm{H}\beta\right)$ as a method of separating the upper and lower branches, with objects with $\log N_2<-0.6$ belonging to the lower branch and objects with $\log N_2>-0.6$ belonging to the upper branch. However, for objects with a redshift $z\gtrsim0.45$, the [\ion{N}{II}] is no longer visible, so this method is not applicable. To separate the higher redshift objects into branches, we have used the OSSY data and analysed how the two branches behave in different plots to obtain a separation criterion. In the left panel of Fig.\ \ref{fig:branch_selection}, where we plot the metallicity obtained with the \citetalias{Pilyugin2016} calibration against $M_*$, we can see that the objects in the lower branch do not exceed $M_*$ values around $\log M_* (M_\odot) \sim 9.3 $. In the middle panel, we plot the parameter $R_{23}$ against the same metallicity to check the regime of the objects as a function of branch. It can be seen that in this plot the objects in the lower branch tend to be in the $R_{23}\gtrsim0.8$ region. On the right we have plotted the two previous variables that we could use to constrain the branches. It is clear that by using a value of $R_{23}>0.85$ and $M_*$ of $\log M_*(M_\odot) < 9.3$ we could separate between the branches objects at higher redshifts where the [\ion{N}{II}] lines are not visible.


\section{Metallicity calibrations}\label{sec:appendixB}
In order to compare the metallicities obtained with different parameterisations, we have performed calibrations between different methods using the OSSY database.

\begin{figure*}[h!]
    \includegraphics[trim= 0cm 0cm 0cm 0cm,width=\textwidth]{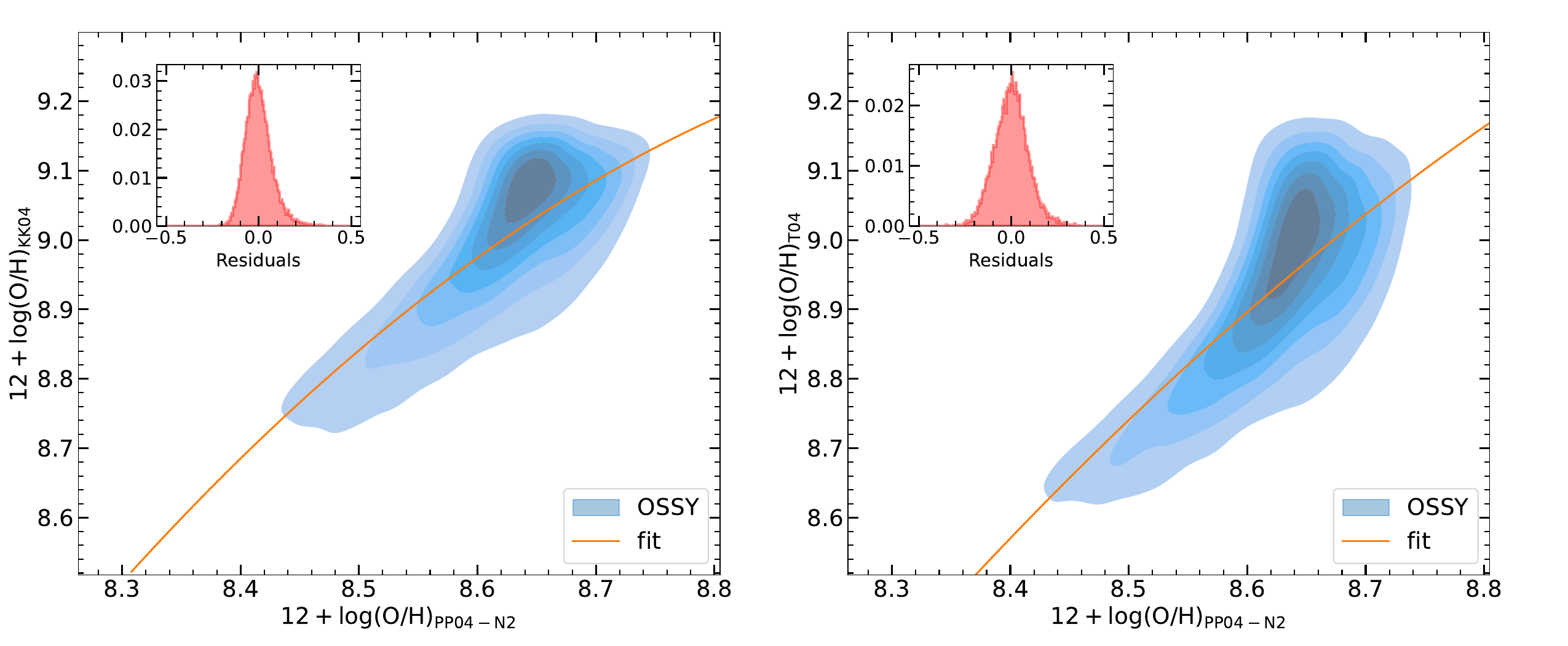}
   \caption{Cross-calibration of metallicity determination methods. On the left, we illustrate the transformation from metallicities following the \cite{Tremonti2004} method to those based on \cite{Kobulnicky2004}. On the right, we depict the conversion from \cite{Kobulnicky2004} metallicities to those derived using the \cite{Tremonti2004} method. The blue circles represent data points sourced from the OSSY catalogue, while the orange line represents a third-order polynomial fit to the OSSY data.} 
    \label{fig:meta_calibrations_N2}
\end{figure*}

The calibration for the transformation from \citetalias{Pettini2004} metallicity, based on the N2 tracer, to the metallicities of \citetalias{Kobulnicky2004} and \citetalias{Tremonti2004} is shown in the Fig. \ref{fig:meta_calibrations_N2}. The calibrations obtained in the fits are:
\begin{equation}
        \left[\meta\right]_{\rm KK04} = -83.91818785 + 20.37854882\times\left[\meta\right]_{\rm PP04-N2} -1.11361199\times\left[\meta\right]_{\rm PP04-N2}^2 
\end{equation}
\begin{equation}
        \left[\meta\right]_{\rm T04} = -58.34509817 + 14.154599852\times\left[\meta\right]_{\rm PP04-N2} -0.7367259\times\left[\meta\right]_{\rm PP04-N2}^2 
\end{equation}

\begin{figure*}[h!]
    \includegraphics[trim= 0cm 0cm 0cm 0cm,width=\textwidth]{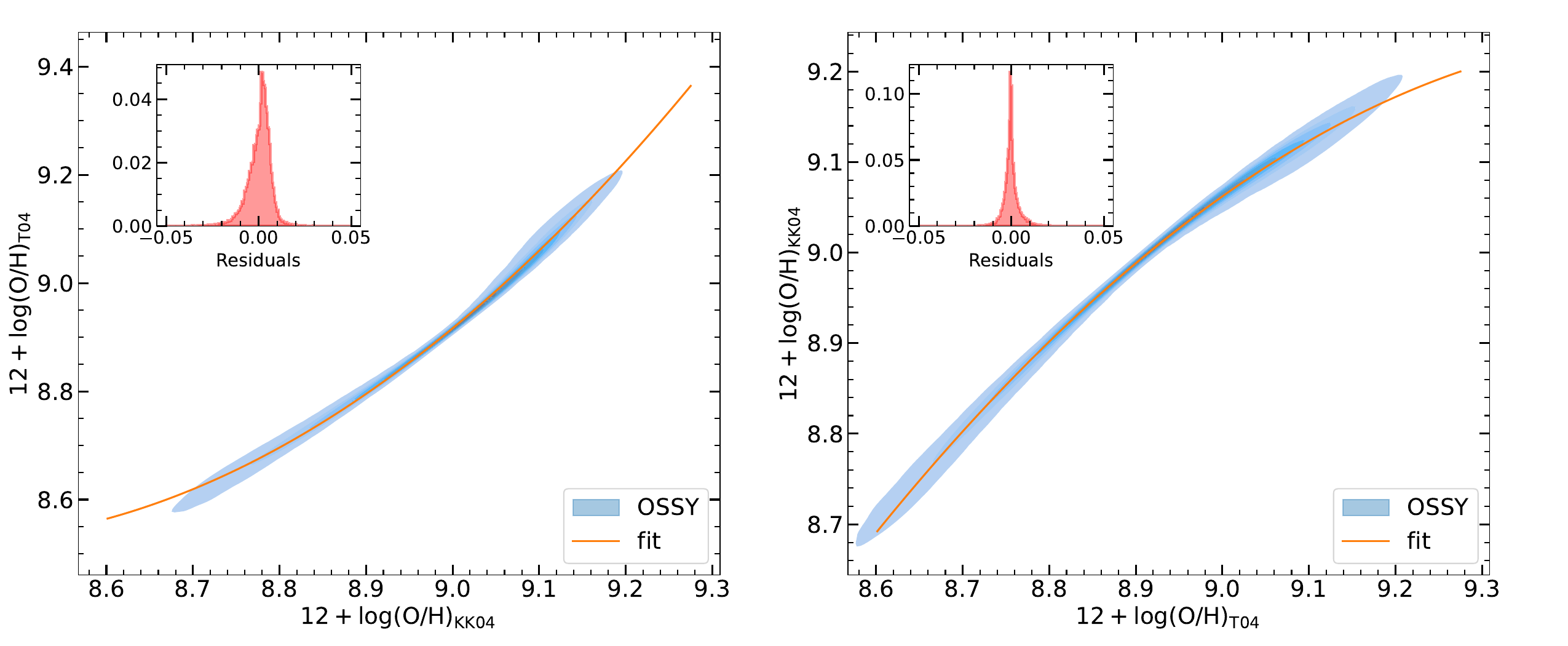}
   \caption{Cross-calibration of metallicity determination methods. On the left, we illustrate the transformation from metallicities following the \cite{Tremonti2004} method to those based on \cite{Kobulnicky2004}. On the right, we depict the conversion from \cite{Kobulnicky2004} metallicities to those derived using the \cite{Tremonti2004} method. The blue circles represent data points sourced from the OSSY catalogue, while the orange line represents a third-order polynomial fit to the OSSY data.} 
    \label{fig:meta_calibrations}
\end{figure*}

The calibration for the transformation from \citetalias{Kobulnicky2004} metallicity to \citetalias{Tremonti2004} metallicity is shown in the left panel of Fig. \ref{fig:meta_calibrations}. The inverse transformation is also shown in the right panel of Fig. \ref{fig:meta_calibrations}. The calibrations obtained in the fits are:
\begin{equation}
        \left[\meta\right]_{\rm T04} = 86.72778038 -18.61508955\times\left[\meta\right]_{\rm KK04} + 1.10771049\times\left[\meta\right]_{\rm KK04}^2
\end{equation}
\begin{equation}
        \left[\meta\right]_{\rm KK04} = -48.25364669 +12.06139695\times\left[\meta\right]_{\rm T04} -0.63255371\times \left[\meta\right]_{\rm T04}^2 
\end{equation}

\end{appendix}
\end{document}